\documentclass[10pt,journal,compsoc]{IEEEtran}

\ifCLASSOPTIONcompsoc
  \usepackage[nocompress]{cite}
\else
  \usepackage{cite}
\fi

\usepackage{enumerate}
\usepackage{graphics} 
\usepackage{epsfig} 
\usepackage{mathptmx} 
\usepackage{times} 
\usepackage{amsmath} 
\usepackage{amssymb}  
\usepackage{comment}
\usepackage{microtype}
\usepackage[dvipsnames]{xcolor}
\usepackage[normalem]{ulem}
\useunder{\uline}{\ul}{}

\definecolor{DarkMagenta}{rgb}{0.5, 0, 0.5}

\usepackage[pdftex, plainpages=false,hypertexnames=true,pdfnewwindow=true,backref=false,colorlinks=true,citecolor=blue,linkcolor=red,urlcolor=blue,filecolor=blue]{hyperref}%
\begin{document}

\title{Electrotactile feedback applications for hand and arm interactions: A systematic review, meta-analysis, and future directions}

\author{Panagiotis~Kourtesis, Ferran~Argelaguet, Sebastian~Vizcay,\\
        Maud~Marchal~\IEEEmembership{Member,~IEEE,}
        and~Claudio~Pacchierotti,~\IEEEmembership{Senior Member,~IEEE,}
\IEEEcompsocitemizethanks{\IEEEcompsocthanksitem P. Kourtesis, F. Argelaguet, and S. Vizcay are with Inria, Univ Rennes, IRISA, CNRS -- 35042 Rennes, France. E-mail: name.surname@inria.fr
\IEEEcompsocthanksitem M. Marchal is with the Univ Rennes, INSA, IRISA, Inria, CNRS -- 35042 Rennes, France and Institut Universitaire de France. \\ E-mail: maud.marchal@irisa.fr
\IEEEcompsocthanksitem C. Pacchierotti is with the CNRS, Univ Rennes, IRISA, Inria -- Rennes 35042, France. E-mail: claudio.pacchierotti@irisa.fr}}

\markboth{
Preprint of Article Submitted to 
IEEE Transactions on Haptics}%
{Shell \MakeLowercase{\textit{emph{et al.}}}: Bare Demo of IEEEtran.cls for Computer Society Journals}

\IEEEtitleabstractindextext{%
\begin{abstract}
Haptic feedback is critical in a broad range of human-machine/computer-interaction applications. However, the high cost and low portability/wearability of haptic devices remain unresolved issues, severely limiting the adoption of this otherwise promising technology. Electrotactile interfaces have the advantage of being more portable and wearable due to their reduced actuators' size, as well as their lower power consumption and manufacturing cost. The applications of electrotactile feedback have been explored in human-computer interaction and human-machine-interaction for facilitating hand-based interactions in applications such as prosthetics, virtual reality, robotic teleoperation, surface haptics, portable devices, and rehabilitation. This paper presents a technological overview of electrotactile feedback, as well a systematic review and meta-analysis of its applications for hand-based interactions. We discuss the different electrotactile systems according to the type of application. We also discuss over a quantitative congregation of the findings, to offer a high-level overview into the state-of-art and suggest future directions. Electrotactile feedback systems showed increased portability/wearability, and they were successful in rendering and/or augmenting most tactile sensations, eliciting perceptual processes, and improving performance in many scenarios. However, knowledge gaps (e.g., embodiment), technical (e.g., recurrent calibration, electrodes' durability) and methodological (e.g., sample size) drawbacks were detected, which should be addressed in future studies.
\end{abstract}

}

\maketitle

\IEEEdisplaynontitleabstractindextext

\IEEEpeerreviewmaketitle

\IEEEraisesectionheading{\section{Introduction}\label{sec:introduction}}

\IEEEPARstart{T}{he} afferent nerves provide the central nervous system with various sensory information, which is essential for perceiving the spatiotemporal continuum (i.e., the world) that surround us. The glabrous skin of the human hand is innervated by 12 diverse types of afferent fibres, which are responsible for perceiving pain, thermal, kinesthetic, and tactile (e.g., form, texture, skin’s motion, motion of exogenous objects, pressure) sensations~\cite{Johnson2000}. This haptic information is considered crucial for facilitating an effective human-machine/computer interaction~\cite{Bach2003,Kaczmarek1991,Pacchierotti2017,Van2009,Pacchierotti2015b}. Indeed, haptic information is needed in human-machine/computer-interaction systems where the user may operate virtual or robotic hands, or simply interact with virtual environments or physical objects in distance. Hence, in a wide range of application scenarios, to enrich interaction fidelity of user's awareness, haptic information is required ~\cite{Rahal2020,Abi2019,Chouvardas2008,Pacchierotti2016,Van2009}.%

Knowing that afferent nerves can be stimulated by electric, mechanical, and thermal stimuli, several haptic interfaces have been developed to elicit such types of haptic feedback~\cite{Kaczmarek1991,Pacchierotti2017,Chouvardas2008}. Mechanical haptic interfaces are the most common types~\cite{Kaczmarek1991}. Specifically, the vibrotactile are those that have been predominantly developed, evolved, and investigated~\cite{Choi2012,Chouvardas2008,Pacchierotti2017,Howard2019,Pacchierotti2015}. While contemporary tactile devices appear to be effective, their cost, portability, and wearability remain issues that should be further addressed~\cite{Choi2012,Pacchierotti2017,Maisto2017}.%

 To this end, electrotactile interfaces have the advantage of being more portable and wearable due to their actuators' size, as well as benefiting from lower power consumption and manufacturing cost, especially when using a high number of actuators/stimulators (e.g., when stimulating multiple areas)~\cite{Bolanowski1988,Chouvardas2008}. Also, its advantageous wearability/portability and lower cost could be further improved by using micromachining means for the fabrication of high-density stimulators~\cite{Chouvardas2008}. 
 
 Electrotactile feedback is provided by a system comprised of electrodes and stimulators (actuators). The electrical current travels through the subdermal area between the anode(s) and cathode(s) and stimulates the nerves endings (i.e., skin's receptors). The area of the skin where the electrode contacts the skin is stimulated, however, the sensation may be spread further when the contact point is near nerve bundles~\cite{Ghafoor2017}. The way electrotactile systems function is therefore completely different from mechanical and thermal tactile interfaces, as electrotactile feedback is not mediated by any skin receptor.%
 
Electrotactile feedback is affected by diverse features of the electrodes (i.e., location, material, and shape), the current (i.e., duration, frequency, and amplitude), and anatomy (i.e., skin’s thickness and location)~\cite{Boldt2014,Hartmann2015,Paredes2015,Strbac2016}. However, issues (e.g., skin irritation, burns, electric shock, and discomfort) have been raised, which indicate that the aforementioned aspects should be carefully selected and adjusted~\cite{Kaczmarek1991}. Nevertheless, nowadays electrodes' materials are designed to mitigate or avoid the aforementioned issues~\cite{Jung2020}, and a calibration of the current features is performed to offer an electrotactile feedback proportionally to the user's perceptual needs (e.g., sensation and discomfort thresholds) and anatomical requirements (e.g., skin's thickness and location)~\cite{Stephens2018}.%

Electrotactile feedback has been successfully implemented in human-computer~\cite{Chouvardas2008,Jung2020} and human-machine interactions~\cite{Bach2003,Stephens2018} for various hand-based applications such as prosthetics~\cite{Stephens2018}, virtual reality~\cite{Jung2020}, robotic teleoperation~\cite{Sagardia2015,Pamungkas2015b}, transparent haptic displays~\cite{Kajimoto2012,Zhao2018}, and augmented haptics~\cite{Groeger2019,Withana2018}. Thanks to the reduced form factor and price, electrotactile feedback is particularly promising for everyday applications, such as gaming and entertainment, as well as Augmented Reality (AR), where users need to interact with real and virtual objects at the same time~\cite{Pacchierotti2017,Maisto2017}. In these areas, the electrotactile feedback is particularly advantageous because it can be delivered by thin, flexible, and transparent electrodes or displays~\cite{Groeger2019,Kajimoto2012,Withana2018,Zhao2018}.%

This paper presents a systematic review and meta-analysis of electrotactile feedback applications for hands and arm interactions. Since the electrotactile applications were more scarce before 2010, and they were discussed by other relevant reviews (e.g., see \cite{Chouvardas2008,Kajimoto2016,Ghafoor2017,Ma2016}, the studies of the last decade (between 2010 and 2021) were considered. The systematic review facilitates a comprehensive presentation of the state-of-art, while the meta-analysis enables a quantitative congregation of the findings, which in turn allow us to offer insights into the state-of-art and suggest future directions.%

Specifically, \autoref{sec:Overview} presents an overview of electrotactile feedback in terms of stimulation of afferent nerves, technical aspects of electrotactile feedback, and type of stimulation (i.e., epidermal and subdermal). \autoref{sec:Applications} introduces the methodology and inclusion criteria of the review and then offers a discussion of the amassed papers clustered by the type of application (i.e., portable devices, augmented haptics, guidance \& notification systems, biomedical devices, and teleoperation \& VR). \autoref{sec:Meta} presents the methods for the meta-analysis, where the frequency of the findings, the sample sizes, and the locations of stimulation are analyzed and discussed. \autoref{ssec:Discussion} offers a comprehensive discussion, where perspectives on the current state-of-art and future directions are discussed in terms of reliability of findings, technical improvements, wearability/portability, and implementations (i.e., prostheses, teleoperation, and VR). Finally, \autoref{sec:Conclusion} provides a concise recapitulation of all the above.%

\section{Overview of Electrotactile Feedback}
\label{sec:Overview}
\begin{figure}[htbp]
 \centering 
 \includegraphics[width=\columnwidth]{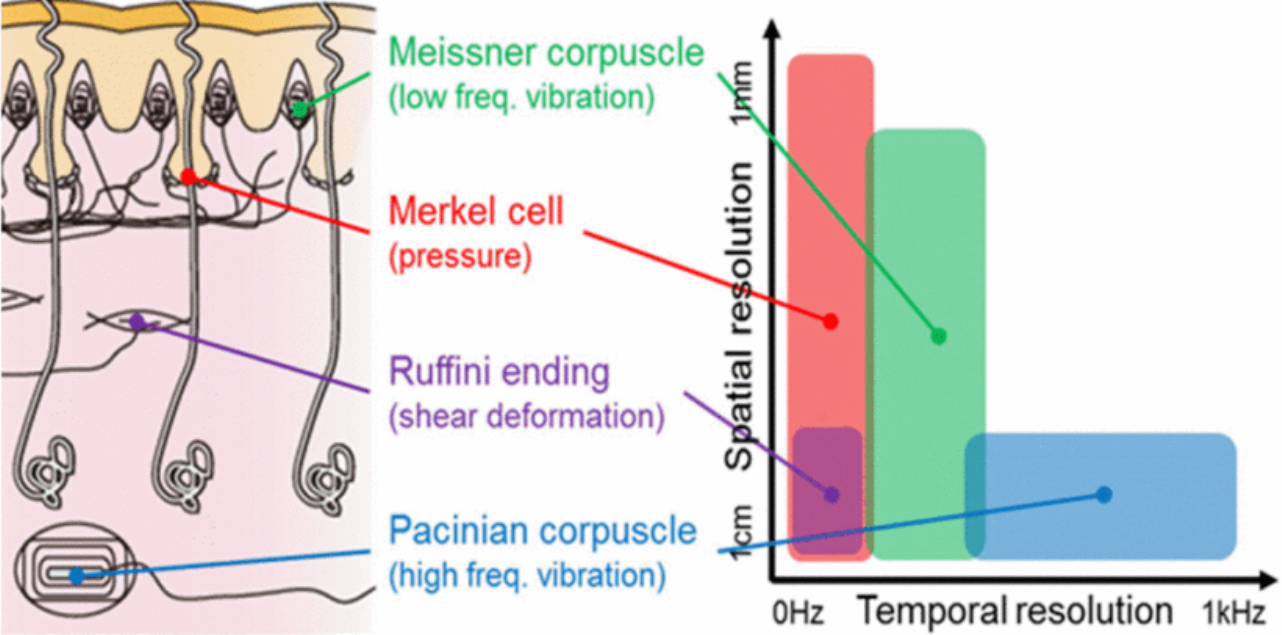}
 \caption{Type of Receptors and their respective Spatial and Temporal Resolution. Image derived from \cite{Yem2017b} }
 \label{fig:RecepterorsType}
\end{figure}

Electrotactile feedback pertains to the electrical current that is produced by a stimulator and travels through the electrodes and the skin layers to stimulate directly the nerve endings of afferent nerves and elicit tactile sensations \cite{Kajimoto2004,He2016}. The afferent nerves that the electrotactile feedback predominantly stimulates to elicit tactile sensations are the mechanoreceptors of the human skin. The four types of mechanoreceptors that the electrotactile feedback predominantly stimulates are the Meissner’s Corpuscles (note that Meissner’s Corpuscles are only present in hairless areas, while hairy areas have hair follicle receptors instead) for eliciting low-frequency vibrational feeling, the Merkel's Cells for rendering pressure and texture, the Pacinian Corpuscle for eliciting high-frequency vibrational feeling, and the Ruffini Endings for mimicking skin stretch\cite{Kajimoto2004}. Each mechanoreceptor has a diverse spatial and temporal resolution (see \autoref{fig:RecepterorsType}) \cite{Kajimoto2004,Yem2017b}. The selective stimulation of the mechanoreceptors, either individually or combined, can potentially elicit diverse tactile sensations \cite{Kajimoto2004,He2016,Yem2017b}.  An overview of the work conducted on the selective electrical stimulation of mechanoreceptors can be found in the survey of Kajimoto \emph{et al.} \cite{Kajimoto2004}, and a quantitative approach of stimulating diverse mechanoreceptors can be explored in the psychophysical experiments of He \emph{et al.} \cite{He2016}.  In this section, we will provide an overview of the electrotactile stimulation, by presenting and discussing the technical features that modulate electrotactile stimulation, as well as the characteristics and applications of epidermal and subdermal stimulation.%

\subsection{Technical Features of Electrotactile Feedback}
\label{ssec:Technical Features}
The sensations that the electrotactile feedback elicits are modulated by several factors such as the electrodes, the skin features, the polarity, and the electrical properties of the pulses (i.e., waveform, amplitude, frequency, and pulse width) \cite{Pamungkas2018,Kajimoto2016}, as well as the calibration process, and the stimulation type. We present below an overview of these technical features of electrotactile feedback (for a comprehensive review on the electrotactile feedback's principles and hardware see Kajimoto \cite{Kajimoto2016}).%

\subsubsection{Electrodes}
\label{sssec:Electrodes}
The material of the electrodes should be taken into account. The electrodes, through an electrochemical process, may produce undesired electrochemical compounds \cite{Brummer1975}. However, electrodes made by noble metals or conductive polymers appear to alleviate or avoid undesired electrochemical reactions \cite{Brummer1975}. Furthermore, the size of the electrodes should also be considered, since small electrodes elicit higher density currents, which cause discomfort to the user \cite{KaczmarekElectrodes}. In contrast, 10mm$^2$ or larger electrodes appear to be more appropriate \cite{KaczmarekElectrodes,Pamungkas2018}. Another aspect is the quantity and density of the electrodes. Solomonow \emph{et al.} studies investigated the distance between electrodes, known as Two-Point Discrimination Threshold (TPDT), where they suggested a TPDT of 7.25mm for the fingertip, 7.73mm for the palm, 8.93mm for the forearm, and 9.48mm for the upper arm (for more see \cite{Solomonow1977, Solomonow1978}). However, more recently, studies have efficiently used a TPDT of 2mm for the fingertips (see \cite{Tsai2019}).%

\subsubsection{Skin Properties}
\label{sssec:Skin}
The thickness of the respective skin area plays an important role in the impedance and sensitivity.  Thicker skin has higher impedance and lower sensitivity to the electrical stimulation \cite{Stephens2018}. Hairless skin (i.e., glabrous skin) occupies approximately 5\% of the human body, and it is thicker than the hairy skin \cite{Pamungkas2018}. Hence, the glabrous skin has higher impedance and lower sensitivity to the electrical stimulation. Furthermore, physiological changes like perspiration (i.e., sweat) may fluctuate skin's conductance that result in an absence of sensation or a discomfort feeling respectively to the fluctuations \cite{Bach2010}. Finally, motion and action may cause deformations (e.g., skin stretching or folding) of the skin that also alter the perceived sensation (e.g., tingling, vibratory, or piercing) and intensity (i.e., absence of sensation or presence of discomfort) of electrotactile feedback \cite{Kajimoto2012b}.%

\subsubsection{Polarity}
\label{sssec:Polarity}
\begin{figure}[htbp]
 \centering 
 \includegraphics[width=\columnwidth]{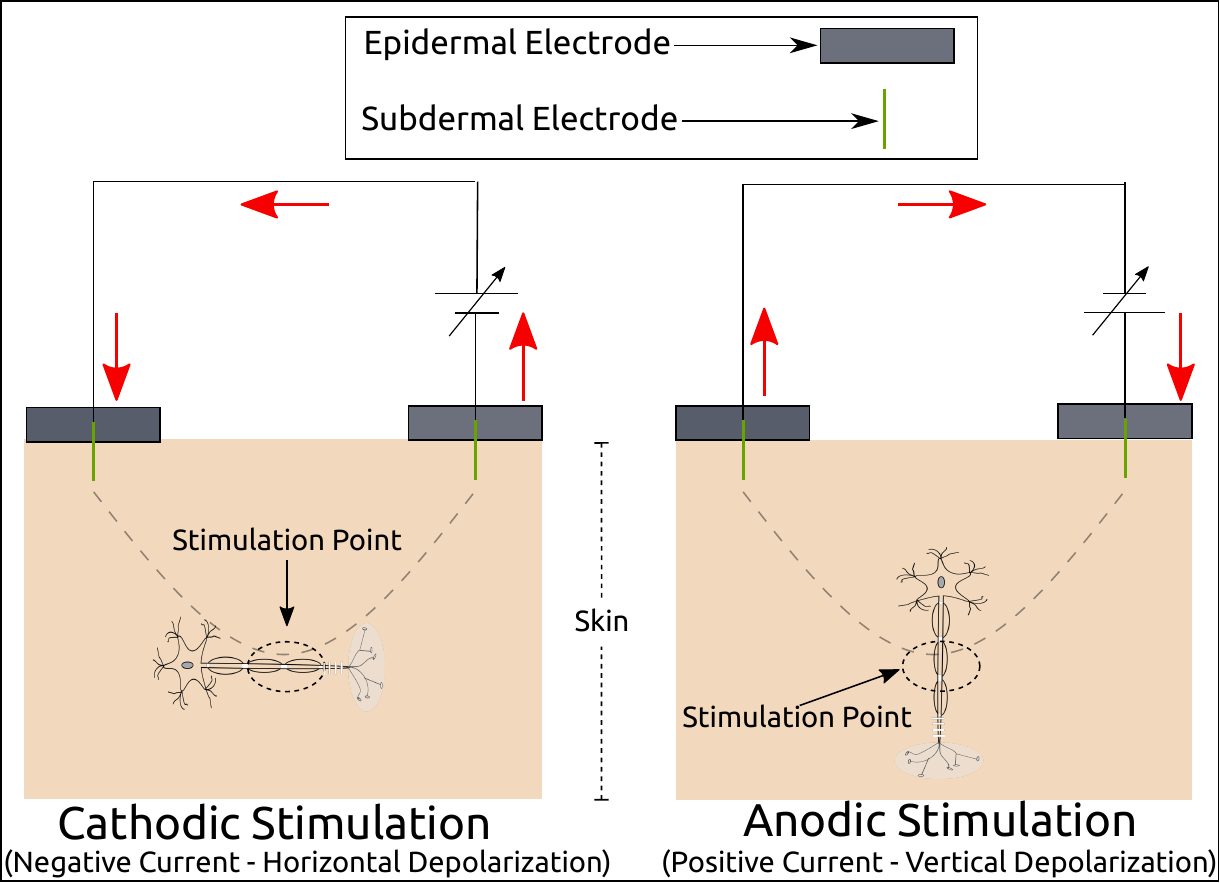}
 \caption{Electrotactile Stimulation: Cathodic Stimulation (Left) and Anodic Stimulation (Right) by either Epidermal or Subdermal Electrodes. The image is an adapted version of Figure 2 in \cite{Yem2017b} }
 \label{fig:ElectroStim}
\end{figure}

The studies of Kaczmarek \emph{et al.} \cite{Kaczmarek1994,Kaczmarek2000} revealed that the positive and negative polarity of electrical current stimulate the human mechanoreceptors in a substantially different way (see \autoref{fig:ElectroStim}). The polarity of the current is respective to the type of stimulation (i.e., Anodic or Cathodic). As displayed in \autoref{fig:ElectroStim}, the anodic stimulation stimulates the receptors via a perpendicular depolarization, while the cathodic stimulation stimulates them through a horizontal (parallel to the skin's surface) depolarization \cite{Yem2017b}. In general, significant differences were identified in the sensation and tolerance of electric current across the human body, where the hand (the glabrous skin area and the fingertips) was found to be more sensitive than other parts like the abdomen or the arm. The negatively charged monophasic current was found to have lower sensation thresholds (i.e., the threshold for perceiving the electrotactile sensation) than the positively charged ones, while these results were not replicated in the fingertips, where negative polarity was found to produce weak and diffuse sensation \cite{Kaczmarek1994}. Furthermore, the positive polarity appears more comfortable than the negative one \cite{Kaczmarek1994,Kaczmarek2000}.%

However, the aforementioned outcomes were pertinent to monophasic stimulation , anodic or cathodic, that only has either a positive or a negative charge respectively . In contrast, the biphasic stimulation that alternates between positive and negative charge appears to be more effective in eliciting diverse tactile sensations \cite{Kaczmarek1992,Pamungkas2018}. Also, biphasic stimulation prevents half-cell reaction, which increases the efficiency of the stimulator, and skin irritation, which augments the comfortability of electrotactile feedback \cite{Kajimoto2002Bi, Pamungkas2018}. The common approach of biphasic stimulation is to provide, in a short time, a negative current subsequently to a positive one, and vice versa, which alleviates the polarization caused by the first stimulation. Given that the depolarization facilitates a seamless stimulation \cite{Kaczmarek1994,Kaczmarek2000}, the advantages of the biphasic approach are of utter importance. Nevertheless, in the last decade, researchers have efficiently used both monophasic (e.g., \cite{Nakamura2014,Kaczmarek2017b,Rahimi2019} and biphasic (e.g.,\cite{Seps2011,Chen2019,Seminara2020}) stimulations.%

\subsubsection{Pulses' Electrical Properties \& Techniques}
\label{sssec:Pulses}
Electrotactile feedback is delivered through electrical pulses. The two main waveforms of the pulses are sinusoidal and squared ones. Due to their ease of implementation and interpretation, the squared pulses are strongly preferred in the applications of electrotactile feedback \cite{Pamungkas2018}. Importantly, the squared pulses were found to facilitate a faster depolarization of the nerves axon, which enhances the effectiveness of the stimulation \cite{PASLUOSTA}. Moreover, the sensation and intensity of electrotactile feedback are modulated by the pulses' amplitude, frequency, and width (a.k.a. pulse width). By modulating one of these parameters or a combination of them, different sensations may be rendered \cite{Aiello,Djozic,Kaczmarek2000,Akhtar2018}. However, in terms of sensation, the relationship between the parameters is non-linear, which is better described with the Weber-Fechner law \cite{Djozic}. In general, the pulse width and amplitude may modulate the perceived intensity, while the frequency may modulate the perceived sensation (e.g., vibratory or tingling) \cite{Aiello,Djozic,Kaczmarek2000,Akhtar2018}.%

The pulses can be delivered distinctly or grouped as bursts. It should be noted that continuous (single) pulses, especially of low intensity and high frequency, may induce an undesired tolerance to electrotactile stimulation \cite{Lindblom}. However, bursts of pulses delivered at lower frequencies (i.e., $<$15Hz) can prevent this tolerance to electrotactile stimulation \cite{Kaczmarek2000}. Nevertheless, a factor that should be considered is the number of pulses per burst, as well as the delay between bursts. Kaczmarek \emph{et al.} \cite{Kaczmarek1992} showed that more than 6 pulses per burst may not produce any improvements in performance. The results of this study showed that bursts embedding 6 pulses with a frequency of 200Hz and a delay per burst of 37ms may improve the performance of electrotactile feedback.%

\subsubsection{Calibration}
\label{sssec:Calibration}
Inappropriate implementation of electrotactile feedback may cause absence of sensation, skin irritation, burns, electric shock, and discomfort ~\cite{Kaczmarek1991}.  A personalized calibration may prevent these undesirable effects by adjusting the intensity and duration of the electrotactile feedback to the personal sensational and discomfort thresholds of the user \cite{Stephens2018}. During the calibration process, the sensation threshold (i.e., the point that the user starts to perceiving/feeling the electrotactile feedback) and the discomfort threshold (i.e., the point that the user starts to perceiving/feeling discomfort) should be defined. These thresholds can be identified by using a step-wise incremental administration of electrotactile feedback. The properties that can be incrementally modulated during the calibration process are the amplitude, frequency, and/or pulse width, respectively to the experimental aims. However, a calibration process should be performed for each stimulation location since the skin's impedance may differ. Also, the calibration process should be recurrent, since the tolerance of the user or the perspiration/moisture levels on the skin may change. A series of models have been created in attempt to optimize and/or automate the calibration process (e.g., \cite{Gregory2009,Fadali2009,Kajimoto2012b,Akhtar2014,Akhtar2018}).%

\subsection{Epidermal \& Subdermal Stimulation}
\label{ssec:Stimulation}
 As discussed in \autoref{sssec:Polarity}, the stimulation may be either anodic or cathodic, where the implementation may be monophasic (i.e., always delivering either an anodic or a cathodic stimulation), or biphasic (i.e., alternating deliverance between anodic and cathodic stimulation). However, the stimulation can be further differentiated by the type of electrodes that were used. The two types are the epidermal (i.e., on the skin's surface) and the subdermal (i.e. penetrating skin's surface) electrodes (see \autoref{fig:ElectroStim}). Thus, epidermal and subdermal stimulations are the two types of stimulation that have been used in the applications of electrotactile feedback for hand and arm interactions. Epidermal stimulation is the approach that has been preferred in the vast majority of the applications, while subdermal stimulation has only been used by a few researchers in the field.  We present below an overview of both stimulation types and their utility.%

\subsubsection{Epidermal Stimulation}
\label{sssec:Epidermal}
The epidermal stimulation refers to electrodes, which are placed on the skin of the user without penetrating it. The electrical current circulates in the subdermal area of the skin between the anodic and cathodic pads (a description of the stimulation is illustrated in \autoref{fig:ElectroStim}). The electrical current stimulates skin’s receptors of this area, which may be expanded to nearby areas when the area includes a nerve bundle(s) \cite{Ghafoor2017}. Epidermal stimulation has been applied on the forearm and upper-arm predominantly for prosthetic hands' users (see below \autoref{ssec:Prostheses}) and on the hand, especially on the fingertips.%

The fingertips are considered the cornerstones of perceiving the surrounding environment, providing the user with haptic information that is essential in various interactions like grasping and manipulating objects \cite{Edin2008,Johnson2000,Pacchierotti2017,Westling1987}. Most haptic interfaces are therefore designed and developed for the fingers and the hands~\cite{Pacchierotti2017,Maisto2017}. In this respect, Kaczmarek \emph{et al.} \cite{Kaczmarek2017} found that users were able to accurately differentiate the frequency and intensity of electrotactile feedback on middle fingertip. Notably, the acceptability of receiving electrotactile feedback on the index fingertip has been examined and confirmed by two user studies with an adequate sample size (e.g., N = 15 or 21)  \cite{Shen2011, Shen2014}. Equally, the sense of touching a surface has been successfully rendered by using electrotactile feedback on the index fingertip \cite{Gregory2011,Kuroda2013,Shen2011}. Similarly, the study of Kuroda and Grondin \cite{Kuroda2013} has effectively rendered the sense of touch elicited by electrotactile feedback on every fingertip, except the pinky’s fingertip.%

Moreover, Sato and Tachi \cite{Sato2010} effectively rendered directional force vectors on the fingertip by activating diverse pads of the electrode in variant intervals. The study offered evidence that spatial and temporal aspects of the stimulation are capable to render force and motion \cite{Sato2010}. In their subsequent study, Sato \emph{et al.} \cite{Sato2011} by rendering diverse combinations of force vectors on the fingertip, managed to render texture properties (i.e., smoothness and roughness). These results were also replicated in other studies (i.e., \cite{Mengoni2011,Peruzzini2012}), which further support the potency of electrotactile feedback to render textural information, and facilitate recognition and discrimination between diverse materials \cite{Peruzzini2012}. Finally, in another follow-up study, Okabe \emph{et al.}, \cite{Okabe2012} by rendering the force vectors, elicited an illusion of multi-directional motion to the users. The users perceived a motion of sliding over a surface by following diverse trajectories.%

Nevertheless, these encouraging results were not limited to the stimulation of the fingers. Another area of interest on the hand, for facilitating human-machine/computer interactions, is the palm \cite{Bach2003,Choi2012,Chouvardas2008,Edin2008,Pacchierotti2017}. However, the skin of the palm folds and stretches in various ways especially in its central area \cite{Son2018}, which may affect the impedance of electrical stimulation. Also, the peripheral regions of the palm have been found to be more receptive to haptic feedback \cite{Son2018}. Since the impedance of electrotactile feedback is an important factor for its effectiveness and pleasantness \cite{Kaczmarek1991}, the electrical stimulation hence should be carefully delivered on the peripheral regions of the palm. In accordance with these methodological suggestions, the users in the study of Boldt \emph{et al.} \cite{Boldt2014} were able to discriminate the spatial (i.e., location) and temporal (i.e., frequency) aspects of the electrical stimulation. Also, recently, Alotaibi \emph{et al.} \cite{Alotaibi2020} conducted a user study (N = 21), where the participants perceived tactile sensations on their palm. The same study also revealed that the amplitude and pulse width of the electrical stimulation on the palm play a significant role in the perceived sensations \cite{Alotaibi2020}.%

\subsubsection{Subdermal Stimulation}
\label{sssec:Subdermal}
As discussed above, skin's conductivity is important for the perceptual effectiveness and pleasantness of the electrotactile feedback. However, the skin's conductivity may fluctuate over time due to physiological changes such as perspiration, which may affect the electrodermal activity \cite{Bach2010}. In response to this drawback, the subdermal stimulation has been counter-proposed for providing the user with stable electrotactile feedback, which may also be more energy-cost-effective \cite{Kitamura2013}.  As \autoref{fig:ElectroStim} displays, subdermal stimulation is similar to the epidermal stimulation, with the only difference that the electrodes in subdermal stimulation are microneedles which penetrate the upper skin's layers. Micro-needle electrodes have been successfully used for providing the user with tactile sensations by stimulating the index fingertip \cite{Kitamura2013} and the forearm \cite{Tezuka2017}. The microneedle electrodes penetrate the glabrous skin and stimulate the deep receptors (e.g., pacini corpuscles) of the finger \cite{Kitamura2015}. Interestingly, this subdermal stimulation was efficient in rendering textural aspects like roughness \cite{Kitamura2015}. However, regarding the stimulation of the forearm, the study of Dong \emph{et al.} \cite{Dong2020} presented discrepant findings, where the subdermal stimulation was found to be significantly less stable (i.e., greater psychophysical variations) than epidermal stimulation. Also, the same study found that epidermal stimulation appears to have better spatial resolution \cite{Dong2020b}. However, subdermal stimulation appears to be advantageous in terms of stability (e.g., compactness and placement) which is important in applications, such as prosthetics, where permanent placement is seen as a more practical  \cite{Dong2020b}. Nevertheless, in human-machine/computer applications, where wearability or portability is more crucial, a non-invasive solution like epidermal stimulation would be more effective and practical.%

\section{Electrotactile Feedback Applications For Hand \& Arm Interactions}%
\label{sec:Applications}
\subsection{Methodology of Systematic Review}
\label{ssec:SystematicReview}
\begin{figure}[h]
 \centering 
 \includegraphics[width= \columnwidth]{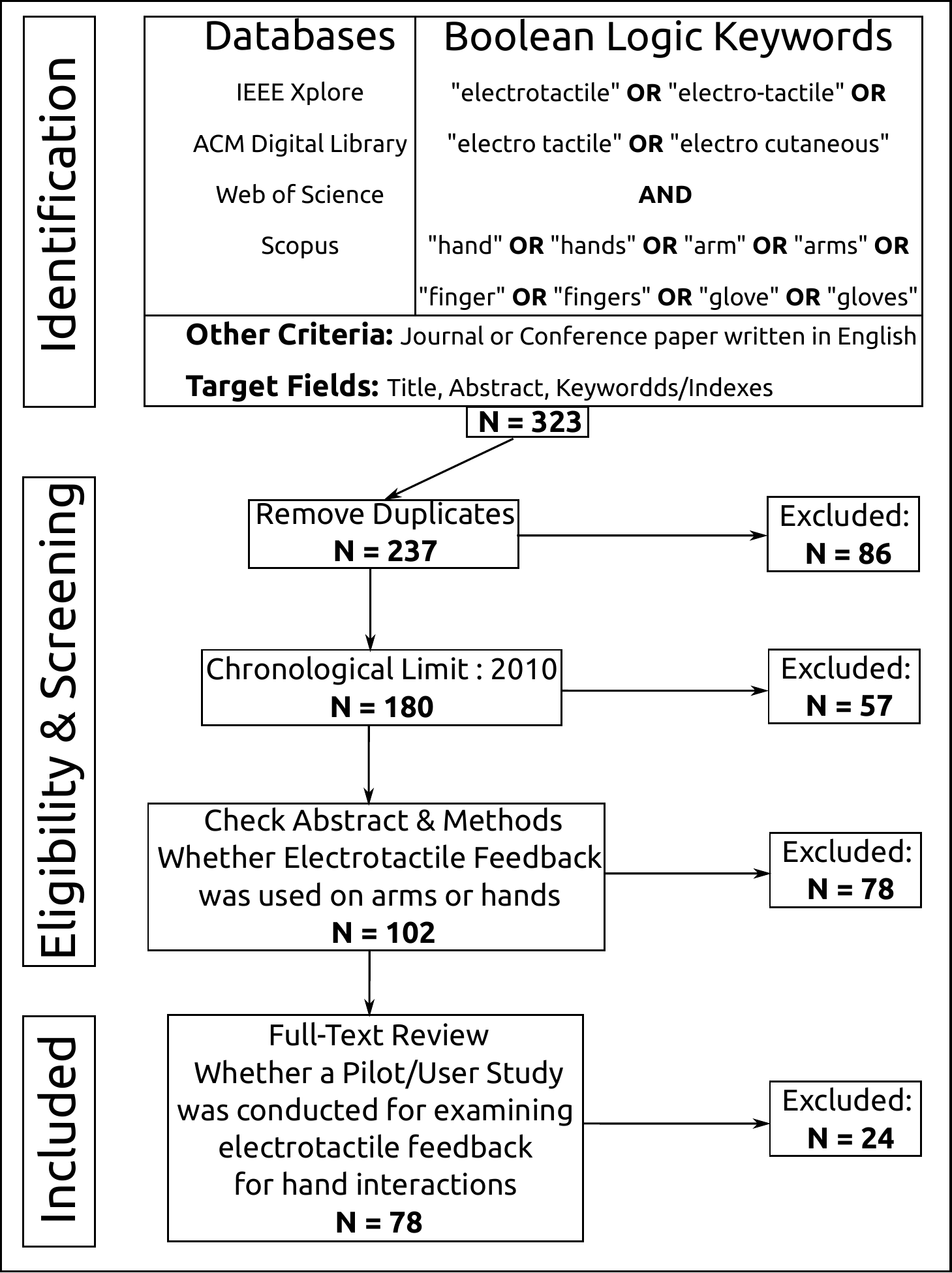}
 \caption{PRISMA Step-Wise Method For Amassing \& Parsing the Studies.}
 \label{fig:PRISMA}
\end{figure}

The Preferred Reporting Items for Systematic Reviews and Meta-Analyses (PRISMA) guidelines were followed \cite{Moher2010}. A step-wise method was hence used for amassing and then filtering the research items based on their relevance to the topic (see \autoref{fig:PRISMA}). Five digital databases were used: \emph{(1) IEEE Xplore Digital Library; (2) ACM Digital Library; (3) ScienceDirect; (4) Web of Science; and (5) Scopus} to ensure a comprehensive collection of electrotactile feedback implementations for hand interactions. Correspondingly, suitable keywords and Boolean logic were used to facilitate an extensive inclusion, yet specificity to the topic (see \autoref{fig:PRISMA}). A chronological limit was set since the technology was predominantly advanced during the last decade. Finally, parsing criteria were used to ensure relevance to the topic (e.g., stimulation on upper limbs; discussion of user study's results). The amassed studies are displayed in \autoref{tab:my-table} clustered by their findings. This section will attempt to provide the reader with an overview of the electrotactile feedback implementations in diverse disciplines for various hand interactions. The amassed studies are differentiated and clustered based on the type of implementation. This categorization allows a discussion specific to the requirements of each type of implementation.%

\subsection{Electrotactile Feedback for Portable Devices}
\label{ssec:Portable}
Another advantage of electrotactile feedback is that it can be easily applied to portable devices. One of the early attempts of this decade is the \emph{Palm Touch Panel} by Fukushima and Kajimoto \cite{Fukushima2011}. This handheld device was at the size of a smartphone, where the front display was the interactive one (i.e., a touchscreen), and the back was the stimulating part, which delivers the electrotactile feedback to the palm. The \emph{Palm Touch Panel} was able to render the sense of touching (i.e., selecting/interacting) by stimulating the corresponding area of the palmar skin \cite{Fukushima2011}. Using a comparable stimulator attached to the back of a smartphone, Khurelbaatar \emph{et al.} stimulated the index fingertip of the user's hand, which was holding the smartphone, while the interaction with the touchscreen was done by the other hand's index \cite{Khurelbaatar2016}. In their study, the users successfully recognized diverse shapes, which indicated the spatial accuracy of the electrotactile feedback \cite{Khurelbaatar2016}.%

Furthermore, on electrotactile touchpads, users who were receiving the electrotactile feedback on the interactive index fingertip reported high acceptability, as well as easiness to differentiate the intensity of stimulation \cite{Tsai2019}. Using a comparable haptic display, the finger tracking was evaluated as highly accurate \cite{Kajimoto2011}. Finally, Yem and Kajimoto \cite{Yem2017} study postulated that the users were perceiving the amount of pressure applied on the haptic interface, by delivering a stimulation proportional to the applied pressure.%

However, the unique advantage of electrotactile feedback is that can be delivered through transparent displays, which are placed in front and/or the back of touchscreen \cite{Altinsoy2011}. In Altinsoy and Merchel \cite{Altinsoy2011} the users received electrotactile feedback on both the hand holding the device (i.e., stimulating the palm) and the index fingertip which was interacting with the touchscreen. Importantly, the users received textural information, which was assisted them with recognizing and differentiating effectively between sandpaper and grooved wood \cite{Altinsoy2011}. Kajimoto \cite{Kajimoto2012} examined the \emph{Skeletouch}, a transparent haptic display, which was used over a smartphone's touchscreen for efficiently rendering the sense of touch by stimulating the index fingertip. Equally, Wang \emph{et al.} \cite{Zhao2018} used \emph{E-Pad}, a transparent electrotactile display over a tablet (see (a) in \autoref{fig:HD}) for rendering motion (i.e., index movement on the touchscreen) and provide guidance (i.e., following the path). The participants were substantially better in following the path when they were receiving electrotactile feedback than when they were receiving vibrotactile feedback \cite{Zhao2018}.%

\begin{table*}[ht]
\Large\addtolength{\tabcolsep}{-100pt}
\caption{Electrotactile Feedback For Hand \& Arm Interactions}
\renewcommand{\arraystretch}{4.5}
\resizebox{515pt}{!}{%
\begin{tabular}{cccccc}
\hline
\fontsize{70}{72}\selectfont\textbf{Findings \& Outcomes} &
\fontsize{70}{72}\selectfont\  \textbf{Studies} &
 \fontsize{70}{72}\selectfont\ \textbf{Sample Size Range} &
 \fontsize{70}{72}\selectfont\ \textbf{Haptic Devices} &
 \fontsize{70}{72}\selectfont\ \textbf{Domain of Application} &
 \fontsize{70}{72}\selectfont\ \textbf{Stimulation Locations} \\ \hline
\\ \fontsize{70}{72}\selectfont\ Acceptability/Suitability &
  \begin{tabular}[c]{@{}c@{}}\fontsize{70}{72}\selectfont\
  \cite{Shen2011,Shen2014,DAlonzo2014,Tezuka2017,Kaczmarek2017,Saleh2018}\\      \fontsize{70}{72}\selectfont\ \cite{Groeger2019,Yang2019,Tsai2019,Rahimi2019,Isakovic2019,Dong2020,Alotaibi2020}\end{tabular} &
  \fontsize{70}{72}\selectfont\ 1 - 21 &
  \begin{tabular}[c]{@{}c@{}} \fontsize{70}{72}\selectfont\ Finger Cap; Free Electrodes; \\      \fontsize{70}{72}\selectfont\ Microneedle Electrodes; \\      \fontsize{70}{72}\selectfont\ Touchpad; Multipad; Armband\end{tabular} &
  \fontsize{70}{72}\selectfont\ None; Prosthetic hand;
  Tablet;   Smartphone &
  \begin{tabular}[c]{@{}c@{}}\fontsize{70}{72}\selectfont\ Index \& Middle Fingertip;   \\   \fontsize{70}{72}\selectfont\    Forearm; Palm\end{tabular} \\
\\ \fontsize{70}{72}\selectfont\ Tactile Sensation &
  \begin{tabular}[c]{@{}c@{}}\fontsize{70}{72}\selectfont\ \cite{Akhtar2014,Alonzo2014,Shi2015,Yem2017,Strbac2019}\\      \fontsize{70}{72}\selectfont\ \cite{Groeger2019,Dong2020,Dong2020b,Alotaibi2020}\end{tabular} &
 \fontsize{70}{72}\selectfont\  1 - 21 &
  \fontsize{70}{72}\selectfont\ Free Electrodes; Touchpad; Armband &
  \fontsize{70}{72}\selectfont\ None; Prosthetic hand &
  \begin{tabular}[c]{@{}c@{}}\fontsize{70}{72}\selectfont\ Index \& Middle Fingertip;   \\   \fontsize{70}{72}\selectfont\    Forearm; Palm\end{tabular} \\
\\\fontsize{70}{72}\selectfont\ Touch Rendering &
  \begin{tabular}[c]{@{}c@{}}\fontsize{70}{72}\selectfont\ \cite{Gregory2011,Fukushima2011,Shen2011,Okabe2012,Kajimoto2012,Huang2012}\\        \fontsize{70}{72}\selectfont\ \cite{Witteveen2012,Wang2013,Kitamura2013,Kuroda2013,Pamungkas2013,Yoshimoto2013}\\      \fontsize{70}{72}\selectfont\ \cite{Geng2014,Hartmann2014,Franceschi2015,Pamungkas2015,Pamungkas2015b,Dadi2018}\end{tabular} &
  \fontsize{70}{72}\selectfont\ 1 - 21 &
  \begin{tabular}[c]{@{}c@{}}\fontsize{70}{72}\selectfont\ Finger Cap; Free   Electrodes;\\  \fontsize{70}{72}\selectfont\     Touchpad; Glove\end{tabular} &
  \begin{tabular}[c]{@{}c@{}}\fontsize{70}{72}\selectfont\ None; Prosthetic hand; Tablet;   \\     \fontsize{70}{72}\selectfont\  Smartphone; Robotic arm; \\      \fontsize{70}{72}\selectfont\ Electronic skin; Virtual Reality\end{tabular} &
  \begin{tabular}[c]{@{}c@{}}\fontsize{70}{72}\selectfont\ All Fingertips; Palm; Dorsal   Hand;\\     \fontsize{70}{72}\selectfont\   Forearm; Upper Arm\end{tabular} \\
\\\fontsize{70}{72}\selectfont\ Texture Rendering &
  \fontsize{70}{72}\selectfont\ \cite{Altinsoy2011,Mengoni2011,Sato2011,Peruzzini2012,Bader2015,Pamungkas2015,Pamungkas2015b,Yoshimoto2015} &
  \fontsize{70}{72}\selectfont\ 1 - 20 &
  \fontsize{70}{72}\selectfont\ Free Electrodes; Touchpad &
  \begin{tabular}[c]{@{}c@{}}\fontsize{70}{72}\selectfont\ None; Tablet; Virtual   Reality\\  \fontsize{70}{72}\selectfont\     Robotic arm; Fingerpad\end{tabular} &
  \begin{tabular}[c]{@{}c@{}}\fontsize{70}{72}\selectfont\ Index Fingertip \& Middle   Phalanx;\\    \fontsize{70}{72}\selectfont\   Palm; Dorsal Hand\end{tabular} \\
\\\fontsize{70}{72}\selectfont\ Motion Perception/Illusion &
  \begin{tabular}[c]{@{}c@{}}\fontsize{70}{72}\selectfont\ \cite{Okabe2012,Damian2012,Franceschi2015,Franceschi2016,Strbac2016}\\      \fontsize{70}{72}\selectfont\ \cite{Xu2016,Franceschi2017,Zhao2018,Achanccaray2020}\end{tabular} &
  \fontsize{70}{72}\selectfont\ 4 - 20 &
  \begin{tabular}[c]{@{}c@{}}\fontsize{70}{72}\selectfont\ Free Electrodes; Touchpad;\\   \fontsize{70}{72}\selectfont\     Multipad; Armband\end{tabular} &
  \begin{tabular}[c]{@{}c@{}}\fontsize{70}{72}\selectfont\ None; Prosthetic hand; \\  \fontsize{70}{72}\selectfont\     Electronic Skin; Tablet; \\   \fontsize{70}{72}\selectfont\    Brain Computer Interface; Virtual Reality\end{tabular} &
  \begin{tabular}[c]{@{}c@{}}\fontsize{70}{72}\selectfont\ Index Fingertip;  \\     \fontsize{70}{72}\selectfont\  Forearm\end{tabular} \\
\\\fontsize{70}{72}\selectfont\ Spatial Perception &
  \begin{tabular}[c]{@{}c@{}}\fontsize{70}{72}\selectfont\ \cite{Kajimoto2011,Geng2014,Boldt2014,Choi2016,   Franceschi2016,Strbac2016,Khurelbaatar2016}\\      \fontsize{70}{72}\selectfont\ \cite{Chai2017,Cheng2017,Franceschi2017,Fares2018,Chai2019,Stanke2020}\end{tabular} &
  \fontsize{70}{72}\selectfont\ 4 - 18 &
  \begin{tabular}[c]{@{}c@{}}\fontsize{70}{72}\selectfont\ Free Electrodes; Touchpad;\\    \fontsize{70}{72}\selectfont\  Multipad; Armband\end{tabular} &
  \begin{tabular}[c]{@{}c@{}}\fontsize{70}{72}\selectfont\ None; Prosthetic hand; \\  \fontsize{70}{72}\selectfont\     Electronic Skin; Internet of Things\end{tabular} &
  \begin{tabular}[c]{@{}c@{}}\fontsize{70}{72}\selectfont\ Index Fingertip; Ring Proximal   Phalanx\\   \fontsize{70}{72}\selectfont\    Forearm; Upper Arm; Wrist\end{tabular} \\
\\\fontsize{70}{72}\selectfont\ Temporal Perception/Discrimination &
  \begin{tabular}[c]{@{}c@{}}\fontsize{70}{72}\selectfont\ \cite{Kuroda2013,Geng2014,Boldt2014,Strbac2016}\\      \fontsize{70}{72}\selectfont\ \cite{Cheng2017, Choi2017, Fares2018,Parsnejad2020,Parsnejad2020b}\end{tabular} &
  \fontsize{70}{72}\selectfont\ 4 - 16 &
  \begin{tabular}[c]{@{}c@{}}\fontsize{70}{72}\selectfont\ Free Electrodes;\\   \fontsize{70}{72}\selectfont\     Multipad; Armband\end{tabular} &
  \begin{tabular}[c]{@{}c@{}}\fontsize{70}{72}\selectfont\ None;Prosthetic hand; \\   \fontsize{70}{72}\selectfont\    Electronic Skin;\\    \fontsize{70}{72}\selectfont\   Internet of Things\end{tabular} &
  \begin{tabular}[c]{@{}c@{}}\fontsize{70}{72}\selectfont\ Index, Middle, Ring, \& Thumb   Fingertips;\\   \fontsize{70}{72}\selectfont\    Forearm; Upper Arm; Palm\end{tabular} \\
\\\fontsize{70}{72}\selectfont\ Pressure   Perception &
  \fontsize{70}{72}\selectfont\ \cite{Xu2014,Choi2017,Yem2017} &
  \fontsize{70}{72}\selectfont\ 5 - 10 &
  \fontsize{70}{72}\selectfont\ Free Electrodes; Touchpad &
  \fontsize{70}{72}\selectfont\ None; Prosthetic hand &
  \begin{tabular}[c]{@{}c@{}}\fontsize{70}{72}\selectfont\ Index \& Middle   Fingertips;\\  \fontsize{70}{72}\selectfont\     Forearm; Upper Arm\end{tabular} \\
\\\fontsize{70}{72}\selectfont\ Stiffness Perception &
 \fontsize{70}{72}\selectfont\  \cite{Xu2014,Cheng2019} &
  \fontsize{70}{72}\selectfont\ 1 - 4 &
  \fontsize{70}{72}\selectfont\ Free Electrodes; Armband &
  \fontsize{70}{72}\selectfont\ None; Prosthetic hand &
  \fontsize{70}{72}\selectfont\ Forearm; Upper Arm \\
\\\fontsize{70}{72}\selectfont\ Roughness Perception &
 \fontsize{70}{72}\selectfont\  \cite{Yoshimoto2015,Kitamura2015,Chai2019} &
  \fontsize{70}{72}\selectfont\ 1 - 15 &
  \fontsize{70}{72}\selectfont\ Free Electrodes;   Microneedle Electrodes &
  \fontsize{70}{72}\selectfont\ None; Prosthetic hand; Fingerpad &
  \begin{tabular}[c]{@{}c@{}}\fontsize{70}{72}\selectfont\ Index Fingertip \& Middle Phalanx; Forearm \end{tabular} \\
\\\fontsize{70}{72}\selectfont\ Slipping Perception &
  \fontsize{70}{72}\selectfont\ \cite{Damian2012,Xu2014} &
  \fontsize{70}{72}\selectfont\ 6 - 9 &
  \fontsize{70}{72}\selectfont\ Free Electrodes &
  \fontsize{70}{72}\selectfont\ None &
  \fontsize{70}{72}\selectfont\ Forearm \\
\\\fontsize{70}{72}\selectfont\  Force   Perception/Control &
  \begin{tabular}[c]{@{}c@{}}\fontsize{70}{72}\selectfont\ \cite{Sato2010,Huang2012,Witteveen2012,Yoshimoto2013,Isakovic2016,Xu2016}\\      \fontsize{70}{72}\selectfont\ \cite{Strbac2017,Chai2019,Li2019,Cheng2019}\end{tabular} &
  \fontsize{70}{72}\selectfont\ 1 - 18 &
  \begin{tabular}[c]{@{}c@{}}\fontsize{70}{72}\selectfont\ Finger Cap; Free Electrodes;   \\ \fontsize{70}{72}\selectfont\  Multipad; Armband\end{tabular} &
  \fontsize{70}{72}\selectfont\ Prosthetic hand; Robotic arm; Virtual Reality &
  \begin{tabular}[c]{@{}c@{}}\fontsize{70}{72}\selectfont\ Index Fingertip;  \\  \fontsize{70}{72}\selectfont\     Forearm; Upper Arm\end{tabular} \\
\\\fontsize{70}{72}\selectfont\  Grasping Performance &
  \begin{tabular}[c]{@{}c@{}}\fontsize{70}{72}\selectfont\ \cite{Witteveen2012,Isakovic2016,Hummel2016,Xu2016}\\      \fontsize{70}{72}\selectfont\ \cite{Strbac2017,Ward2018,Li2019,Cheng2019,Achanccaray2020}\end{tabular} &
  \fontsize{70}{72}\selectfont\ 1 - 20 &
  \begin{tabular}[c]{@{}c@{}}\fontsize{70}{72}\selectfont\ Free Electrodes; Multipad;   \\  \fontsize{70}{72}\selectfont\     Armband; Glove\end{tabular} &
  \begin{tabular}[c]{@{}c@{}}\fontsize{70}{72}\selectfont\ Prosthetic hand; Robotic arm;   \\  \fontsize{70}{72}\selectfont\     None; Brain Computer Interface; Virtual Reality\end{tabular} &
  \begin{tabular}[c]{@{}c@{}}\fontsize{70}{72}\selectfont\ All Fingertips;\\   \fontsize{70}{72}\selectfont\     Forearm; Upper Arm\end{tabular} \\
\\\fontsize{70}{72}\selectfont\ Augmented Haptics &
  \fontsize{70}{72}\selectfont\ \cite{Yoshimoto2013,Yoshimoto2015,Yoshimoto2016,Withana2018} &
  \fontsize{70}{72}\selectfont\ 1 - 12 &
  \fontsize{70}{72}\selectfont\ Free Electrodes;   Wristband &
  \fontsize{70}{72}\selectfont\ None; Fingerpad; Virtual Reality &
  \begin{tabular}[c]{@{}c@{}}\fontsize{70}{72}\selectfont\ Index Fingertip \& Middle   Phalanx; \\  \fontsize{70}{72}\selectfont\     Thumb Middle Phalanx\end{tabular} \\
\\\fontsize{70}{72}\selectfont\ Effective Guidance &
  \fontsize{70}{72}\selectfont\ \cite{Yoshimoto2016,Zhao2018} &
  \fontsize{70}{72}\selectfont\ 12 &
  \fontsize{70}{72}\selectfont\ Free Electrodes; Touchpad &
  \fontsize{70}{72}\selectfont\ None; Tablet &
  \begin{tabular}[c]{@{}c@{}}\fontsize{70}{72}\selectfont\ Index Fingertip \& Middle   Phalanx; \\   \fontsize{70}{72}\selectfont\    Thumb Middle Phalanx\end{tabular} \\
\\\fontsize{70}{72}\selectfont\ Validated Teleoperation &
  \fontsize{70}{72}\selectfont\ \cite{Sagardia2015} &
  \fontsize{70}{72}\selectfont\ 5 &
  \fontsize{70}{72}\selectfont\ Glove &
  \fontsize{70}{72}\selectfont\ Robotic Arm; Virtual Reality &
  \fontsize{70}{72}\selectfont\ All Fingertips \\
\\\fontsize{70}{72}\selectfont\ Successful Braille Text   Reading &
  \fontsize{70}{72}\selectfont\ \cite{Liu2016,Rahimi2019b} &
 \fontsize{70}{72}\selectfont\  1 - 4 &
  \fontsize{70}{72}\selectfont\ Touchpad &
  \fontsize{70}{72}\selectfont\ Camera; (Braille) Text Reading Assistive System &
  \fontsize{70}{72}\selectfont\ Index Fingertip \\
\\\fontsize{70}{72}\selectfont\  Speech Reception &
  \fontsize{70}{72}\selectfont\ \cite{Huang2017,Huang2017b} &
  \fontsize{70}{72}\selectfont\ 10 - 14 &
  \fontsize{70}{72}\selectfont\ Tactile Transducer &
  \fontsize{70}{72}\selectfont\ Cochlear Implant &
  \fontsize{70}{72}\selectfont\ Index Fingertip \\ 
\\\fontsize{70}{72}\selectfont\ Successful Virtual Typing &
 \fontsize{70}{72}\selectfont\  \cite{Pamungkas2019} &
  \fontsize{70}{72}\selectfont\ 6 &
 \fontsize{70}{72}\selectfont\  Free Electrodes &
 \fontsize{70}{72}\selectfont\  None &
  \fontsize{70}{72}\selectfont\ All Proximal Phalanxes \\ \hline
\\\end{tabular}%
\renewcommand{\arraystretch}{2}
}
\label{tab:my-table}
\end{table*}
\subsection{Augmented Haptics}
\label{ssec:Augmented}
As seen above, the electrotactile feedback can be delivered over other surfaces (i.e., tablets and smartphones). This also suggests electrotactile feedback's potentiality for augmenting haptic information. Indeed, Yoshimoto \emph{et al.} \cite{Yoshimoto2013} by stimulating the middle phalanx of the index, were able to provide the users with a sense of touch on the fingertip, while the fingertip was at mid-air point over the target-object. In addition, the users were able to perceive the pressing force on the augmented surface of the target-object \cite{Yoshimoto2013}. Comparably, other studies were able to render textural information by stimulating the index fingertip \cite{Bader2015} or middle phalanx \cite{Yoshimoto2015}. Interestingly, the augmented haptic information, by stimulating the middle phalanxes of index and thumb, was implemented for guiding the user's hand movements and significantly improving the performance on a carving task \cite{Yoshimoto2016}.%

However, the above studies have used "traditional" electrodes, while one of the benefits of electrotactile feedback is that can be offered through thin, flexible, and/or transparent electrodes. Withan \emph{et al.} \cite{Withana2018} implemented thin, flexible, and transparent electrodes, which were also connected to a relatively small stimulator (see (b) in \autoref{fig:HD}). This electrotactile device not only presented desirable wearability and portability, but it was also capable of delivering augmented tactile sensations of various surfaces (e.g., a toy car, pencil, paper, and human skin) and while performing diverse tasks in physical (e.g., writing) and virtual reality (i.e., interacting with virtual objects) \cite{Withana2018}. Lastly, in a follow-up study, Groeger \emph{et al.} \cite{Groeger2019} used similar 3-D printed electrodes connected to a portable controller (stimulator), which they then were attached on the surface various objects (e.g., pen or smartphone's case) instead of the human skin (e.g., on the fingertip). This study demonstrated the acceptability of the electrotactile feedback, and an effective augmentation of tactile sensations \cite{Groeger2019}. Although these multi-purpose electrodes can be easily 3-D printed \cite{Groeger2019}, an important drawback is that their durability is limited to up to 8 hours in an office environment\cite{Withana2018}.%

\subsection{Guidance \& Notification}
\label{ssec:GN}
As discussed above, the study of Yoshimoto \emph{et al.} \cite{Yoshimoto2016} showed that electrotactile feedback may be efficiently used for providing guidance to the user for performing tasks (i.e., carving), which require fine movements. In the same direction, the study of Pamungkas and Turnip \cite{Pamungkas2019} examined how the electrotactile feedback may assist the user with virtual typing (i.e, mid-air interactions with a virtual keyboard). By stimulating the lower phalanxes of all five fingers, their study showed that electrotactile feedback facilitates a substantially faster learning curve for virtual typing \cite{Pamungkas2019}. The guidance in both studies is provided by notifying the users that their movements deviate from the optimal ones.%

Hence, electrotactile feedback may also be used as a medium for delivering notifications. Indeed, electrotactile feedback has been found to be effective for facilitating communication between the user and Internet of Things (IoT) devices~\cite{Parsnejad2020}. Notably, Stanke \emph{et al.} \cite{Stanke2020} compared the efficiency of electrotactile and vibrotactile feedback in providing notifications to IoT users by stimulating the left hand's wrist and ring proximal phalanx. Their results indicated that, compared to vibrotactile feedback, electrotactile feedback has a substantially better spatial resolution and facilitates a significantly improved recognition of notification patterns~\cite{Stanke2020}. Contemporary wearable devices like the \emph{Apple Watch}, which are worn on the wrist, provide predominantly vibrotactile feedback to the user~\cite{Pacchierotti2017}. Thus, the aforementioned results indicate that electrotactile feedback may be an alternative with improved performance, while it also alleviates financial burden due to its cost-effective design and fabrication.%

\subsection{Biomedical Applications: Assistive Devices}
\label{ssec:BioAssistive}
Electrotactile Feedback has also been implemented for biomedical applications. As discussed above, electrotactile feedback has been effectively delivered on the index fingertip. Likewise, researchers have administered it on the index fingertip in order to provide the users with a sensory substitution (e.g., haptic-visual or haptic-audio), which may improve the everyday lives of individuals with respective sensory impairments.%

\begin{figure*}[ht]
 \centering
 \includegraphics[width=\textwidth, height= 400pt]{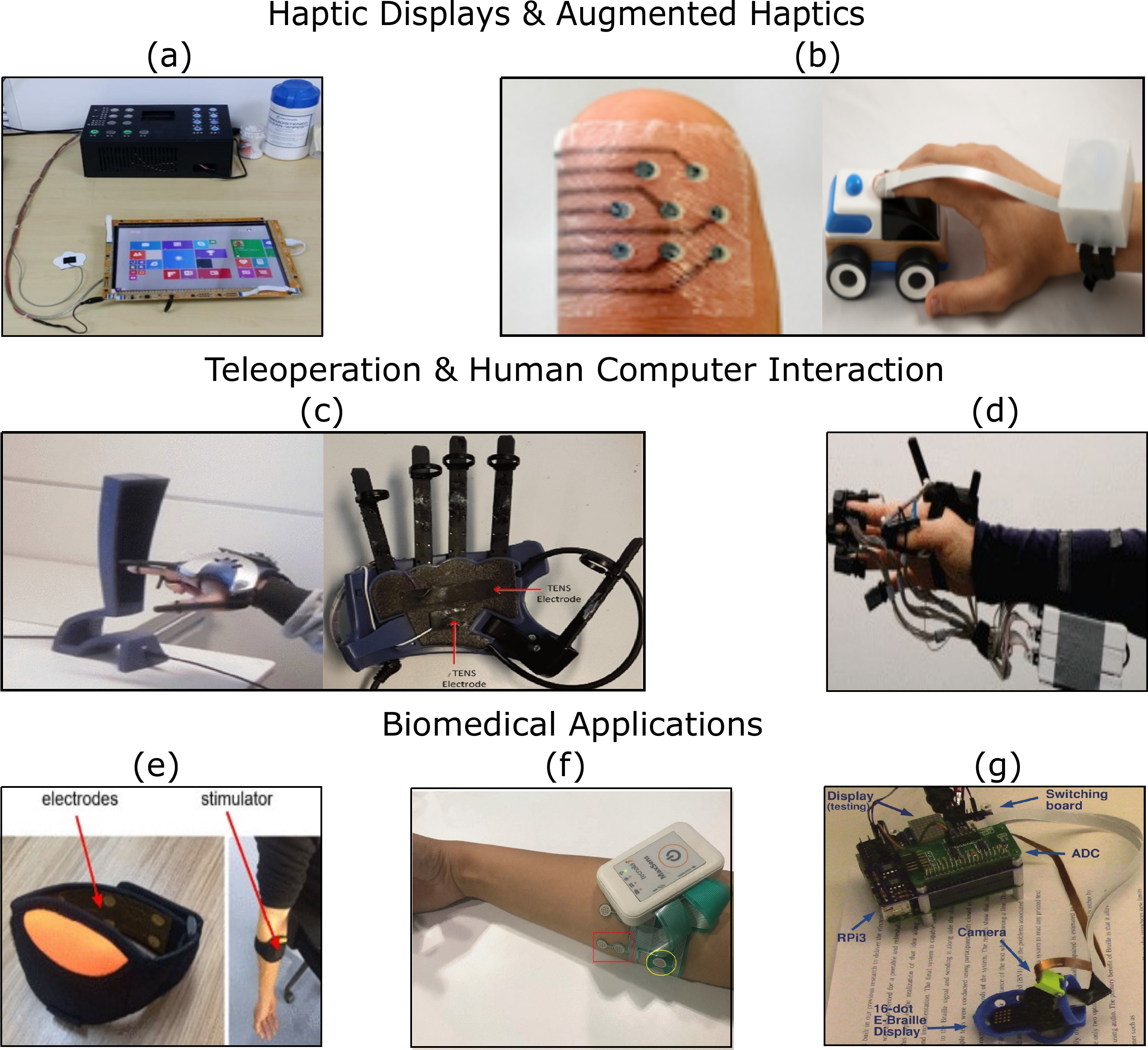}
 \caption{Representative examples of electrotactile feedback systems per implementation:
 (a) transparent touchpad \cite{Zhao2018}; (b) thin epidermal electrodes for augmented haptics \cite{Withana2018};
 (c) glove for dorsal hand stimulation \cite{Pamungkas2015b}; (d) multi-finger stimulation \cite{Hummel2016}; (e) armband for prosthetics \cite{Cheng2017}; (f)  armband for prosthetics \cite{Strbac2019}; (g) braille reading system \cite{Rahimi2019b}.}%
 \label{fig:HD}
\end{figure*}

For substituting visual with haptic information, Liu \emph{et al.} \cite{Liu2016} developed the \emph{Finger Eye}, a wearable text reading assistive system for the blind and visually impaired individuals. In their pilot study, they successfully converted visual data (i.e, text displayed on a screen or paper), which were captured by a Pan-Tilt-Zoom (PTZ) camera, into haptic information correspondingly to the braille reading system. Hence, using text recognition, Liu \emph{et al.} \cite{Liu2016} actually translated conventional text into braille. However, a disadvantage was that the camera was bulky (i.e., normal size of a PTZ camera) which prevents the wearability and portability of such text reading assistive system. Nevertheless, Rahimi \emph{et al.} \cite{Rahimi2019b} developed a significantly more portable and wearable similar device (see (g) in \autoref{fig:HD}). Also, Rahimi et al, \cite{Rahimi2019,Rahimi2019b} attempted to provide an automated recurrent calibration process. Although more research is needed towards this direction, this was a step towards achieving an automated regulation of stimulation power, which will address one of the main drawbacks of electrotactile feedback.%

Furthermore, Huang \emph{et al.} \cite{Huang2017,Huang2017b} substituted audio with haptic information. Through a tactile transducer, they administered electrotactile stimulation on the index fingertip to ameliorate speech recognition of cochlear implant users (i.e., with hearing impairment) \cite{Huang2017,Huang2017b}. They converted speech sound amplitude and frequency into an electric signal with equivalent properties. However, the frequency of the electrical signal did not overpass the 200Hz to ensure temporal discrimination of the amplified speech. Indeed, cochlear implant users showed a significantly improved speech recognition in noise \cite{Huang2017,Huang2017b}. The significantly improved speech recognition was observed both in Mandarin \cite{Huang2017} and English speakers \cite{Huang2017b}.%

\subsection{Biomedical Applications: Prostheses}
\label{ssec:Prostheses}
Somatosensory information plays a crucial role in both motor performance as well as embodied cognition, which makes it vital for prosthetic hands/arms users \cite{Stephens2018,Sensinger2020}. Notably, the sense of embodiment has been suggested to increase the acceptance of a prosthetic hand/arm \cite{Stephens2018,Sensinger2020}. Electrotactile, vibrotactile, or combined feedback have been found to increase both motor performance of prosthetic hand/arm users, as well as the sense of embodiment, although more research is required to further explore those advantages \cite{Stephens2018,Sensinger2020}. Regarding electrotactile feedback, the majority of the implementations are related to prostheses of upper limbs, which will be discussed here.%
 
The rationale is that the electrical stimulation is provided to the forearm or the upper arm, by electrodes placed in the circumference of the limb (a.k.a. non-somatotopic feedback), where the stimulation of each electrode attempts to represent the tactile sensation of a hand's part (e.g, a finger and/or the palm.). Indeed, the users can differentiate each stimulation and associate it correctly with the corresponding finger \cite{Choi2016}, without requiring somatotopic feedback (i.e., the electrodes to be placed near or at the amputation area) \cite{Chai2017}. Also, in the hybrid systems (i.e., vibrotactile and electrotactile), the electrotactile feedback can be easily differentiated from the vibrotactile feedback \cite{DAlonzo2014}. Notably, the electrotactile feedback has been well accepted by the users \cite{Isakovic2019}. Given that the somatosensory information is provided by the electrotactile and the motor abilities by electromyography (EMG) means (i.e., measuring the muscle movement of the forearm, informs on applying the corresponding hand movement and force), a synchronization between electrotactile feedback and EMG is paramount. Strbac \emph{et al.} \cite{Strbac2019} study achieved an effective synchronization between electrotactile feedback (see (f) in \autoref{fig:HD} for the electrotactile device) and EMG, while the users were able to perceive tactile sensations. However, a drawback of electrotactile feedback is that it necessitates recurrent calibration for providing personalized and pleasant feedback. Isakovic \emph{et al.} \cite{Isakovic2019} offered an optimized calibration process, though follow-up research is required.%

Furthermore, the sense of touch has been repeatedly rendered with success by delivering electrotactile feedback \cite{Geng2014,Alonzo2014,Hartmann2014,Franceschi2015}. However, like the vast majority of the relevant studies, the electrotactile feedback was administered on the forearm. Nonetheless, the study of Wang \emph{et al.} \cite{Wang2013} replicated the successful touch rendering by stimulating the upper arm. Also, considering that spatial aspects are crucial for hand interactions, the location of electrotactile feedback has been found that can be easily differentiated \cite{Chai2017,Fares2018}. Beyond the position, the orientation and direction of the stimulation were successfully identified by the user \cite{Franceschi2015}. Importantly, the study of Geng and Jensen \cite{Geng2014} revealed a  spatiotemporal identification of the artificial finger's position and the pulses' width, where the users recognized and paired the current position of the active finger, and/with the duration of stimulation. These results thus support the necessary perceptual mechanisms for hand interactions.%
  
However, additional perceptual processes are required for hand interactions, which pertain to shape, size, textural, and motion features.  In this direction, the earliest attempt in this decade was made by the study of Damian \emph{et al.} \cite{Damian2012}, where an electrotactile feedback on the forearm allowed the users to perceive the motion of the grabbed object, as well as the slippery and the grip force. Similarly, Franceschi \emph{et al.} \cite{Franceschi2017} showed that the users can recognize the shape, trajectory, and direction of a moving target-object. Also, the participants in Choi \emph{et al.} \cite{Choi2017} study received stimulation on the upper arm, which enabled them to perceive the pressure that the prosthesis applied on the surface. Equally, the participants of Huang \emph{et al.} \cite{Huang2012} by receiving a stimulation on the forearm, were able to perceive the applied force. Finally, Chai \emph{et al.} \cite{Chai2019} showed that the electrotactile feedback on the forearm facilitates discrimination of the size and texture (i.e., hardness and softness) of the target object, and it also assists with differentiation of the applied grasping force. Consequently, these results indicate the potency of electrotactile feedback to improve the grasping and handling performance of prostheses' users. Nevertheless, these benefits should be studied directly, especially in patients (i.e., amputees), since the aforementioned studies majorly recruited abled (i.e., healthy) individuals.%
 
Several studies have indeed examined the benefits of electotactile feedback (predominantly applied on the forearm) on hand interactions (e.g., grasping) in amputees. The study of Strbac \emph{et al.} \cite{Strbac2016} found that the set up of the electrotactile system on amputees was effortless, and the amputees were able to perceive kinesthetic information such as aperture, grasping force, and wrist rotation. Also, Isakovic \emph{et al.} \cite{Isakovic2016} demonstrated that electrotactile feedback assisted the amputees to have a significantly improved force perception, as well as a substantially improved grasping accuracy and precision. Substantial improvements in force perception and grasping performance of amputees were also found in the study of Witteveen \emph{et al.} \cite{Witteveen2012}, where the users were controlling a virtual hand; and in the study of Strbac \emph{et al.} \cite{Strbac2017}, where the users were controlling a robotic hand (\emph{Michelangelo hand}). Notably, the amputees in the study of Xu \emph{et al.} \cite{Xu2016}, beyond a significantly improved grasping and force control, showed also significant improvements in grasping stability and movement speed. These results were also replicated in the study of Cheng \emph{et al.} \cite{Cheng2019}, by using a wearable electrotactile system (see (e) in \autoref{fig:HD}) for stimulating the upper arm, they also observed that the individuals with amputations showed significant improvements in differentiating materials with diverse level of stiffness. Of note, in contrast with the above-mentioned studies, which implemented a closed-loop between EMG and electrotactile feedback, the study of Achanccaray and Hayashibe \cite{Achanccaray2020} implemented the electrotactile feedback in combination with a brain-computer-interface (BCI), where the latter was controlling the prosthetic's movements and actions. Comparably to the closed-loop systems, their study revealed that the conjunction of BCI and electrotactile feedback facilitates a substantially improved motor imagery, and significant improvements in flexion and extension, which are the essential grasping movements \cite{Achanccaray2020}.%

\subsection{Teleoperation \& VR}
\label{ssec:TeleVR}
Observing the advantages the electrotactile feedback brings to the control of prosthetic hands, one would reasonably expect these advantages would also apply to the teleoperation of robotic hands and other mechatronics. However, a significant difference between prosthetic and robotic-teleoperation is that the feedback is non-somatotopic (i.e., the feedback is applied on an anatomical part which is in-distance from the moving/interactive anatomical part) while in the latter is somatotopic (i.e., the feedback is applied on the corresponding anatomical part which initiates the actions). Hence, in teleoperation, the user expects to receive haptic feedback on the same area of the hand with the corresponding area of the robotic hand which makes the contact with the target object. Nevertheless, despite the essential differences, the studies which have implemented electrotactile feedback for teleoperation purposes have produced findings comparable to the findings on prostheses.%

Pamungkas and Ward \cite{Pamungkas2013} developed a glove that applies electrotactile feedback on the dorsal area of the hand (see (c) in \autoref{fig:HD}), which was implemented for teleoperating a robotic hand. In their pilot study, they rendered the sense of touch, which allowed the operator to successfully place a peg in a hole~\cite{Pamungkas2013}. In a subsequent pilot study, they tele-operated a robotic drill through immersive VR, where the operator performed drilling with dexterity and accuracy~\cite{Pamungkas2015b}. In Pamungkas and Ward \cite{Pamungkas2016} pilot study, they used the same glove to convey the sensation of contacts in immersive VR for diverse use cases such as playing with a virtual bouncing ball, setting up a campfire, feeling different surface textures, and shooting a gun. 

 Notably, the study of Yem and Kajimoto \cite{Yem2017b}, using a device called \emph{“Finger Glove for Augmented Reality” (FinGAR)}, rendered macro roughness, friction, fine roughness, and
hardness in VR. Their study (N = 10) showed that cathodic stimulations combined with skin deformation modulate macro roughness and hardness, while high-frequency vibration combined with anodic stimulation modulate friction and fine roughness. In another study, Yem and collaborators \cite{Yem2018} showed that electrotactile feedback could inform on the directions of the illusory force sensation (flexion and extension), and render tactile sensations such as the softness, hardness and stickiness of a virtual object. Comparably, in the study (N = 8) of Vizcay et al. \cite{Vizcay2022}, 6 tactile effects were rendered for providing directional (finger's movement), pressure (dynamic press), and contact (tapping) information in VR.  While these are promising results, further research is required in studies with adequately large sample sizes.%

On the other hand, Sagardia \emph{et al.} \cite{Sagardia2015} developed a VR teleoperation system integrated with and an electrotactile delivery system composed of a small tactor with eight electrodes for each finger (see (d) in \autoref{fig:HD}), in an attempt to offer a simulation/training for on-orbit servicing (OOS) missions. The motor movement was controlled by a bimanual haptic device HUG (i.e., light-weight robot arms) and the electrotactile feedback was delivered to each fingertip. Their system was tested in three VR assembly tasks for remote OOS missions, which included pressing a button, switching a lever switch, and pulling a module from its slot. The VR teleoperation training showed comparable results to the physical training, while the users reported that the VR training was highly realistic~\cite{Sagardia2015}. Importantly, their results were replicated and their system was further scrutinized in a user study with an adequate sample size (N = 19)~\cite{Hummel2016}. In this follow-up study of Hummel \emph{et al.} \cite{Hummel2016} the significant difference was that the motor movements in VR were not facilitated by bimanual robotic hands, instead, the hands of the users were tracked, hence their hand-movements were directly mirrored in VR. Their user study found that the electrotactile feedback substantially improved the grasping performance of the users, while it significantly decreased their workload~\cite{Hummel2016}. These results hence appear to replicate the benefits of electrotactile feedback that were seen in prostheses (e.g., improved grasping), and indicate that similar advantages can be seen in VR for computer- and/or machine-human interactions.%
\section{Meta-Analysis of Applications \& Studies}
\label{sec:Meta}
In order to better analyze and categorize the above works, we carried out a meta-analysis of their main characteristics, extracting and converting them into variables. We considered:
\begin{enumerate}[(i)]
        \item \emph{Findings}: 
        \begin{enumerate}[(a)]
        \item The frequency of the findings pertaining to the conclusions of each study are considered. The findings are analyzed by a natural language processing (NLP) approach to indicate the most frequent findings/conclusions.
        \item The field of application is also considered, including biomedical engineering, computer science, robotics, mechanical engineering, and psychology.
        \item The rendered information, sense or illusion, such as tactile, kinesthetic, thermal, and guidance.
        \end{enumerate}
        \item \emph{Sample Size}: \begin{enumerate}[(a)]
        \item The sample size is divided into clusters (i.e., N $\leq$ 10, N $\geq$ 10 , N $\geq$ 15, and N $\geq$ 20) to evaluate the frequency of each sample size cluster.
        \item The type of study is also defined by the sample size. The studies are dichotomized to pilot (N $\leq$ 10) and user studies (N $>$ 10).
        \end{enumerate}
        \item \emph{Stimulation Locations and Devices}:
        \begin{enumerate}[(a)]
        \item The stimulation's locations, including the forearm, the upper arm, the wrist, the dorsal hand, the palm, the phalanxes, and the fingertips.
        \item The type of electrotactile device, categorized into free electrodes (i.e., only the electrodes are on the skin), multipad (multiple electrodes on a pad), armband (i.e., electrodes and stimulator are on the arm), microneedle electrodes, finger cap, wristband (same as armband but on the wrist), touchpad, and glove.
        \item The connected device, which includes the other devices or systems that the haptic interfaces are connected with (e.g., prosthetic hand, tablets, smartphones, tablets, cochlear implants, VR, brain-computer interfaces,  and internet of things). 
        \end{enumerate}
\end{enumerate}%
\subsection{Findings' Frequency}
\label{ssec:Findings}
\begin{figure}[htbp]
 \centering 
 \includegraphics[width=\columnwidth]{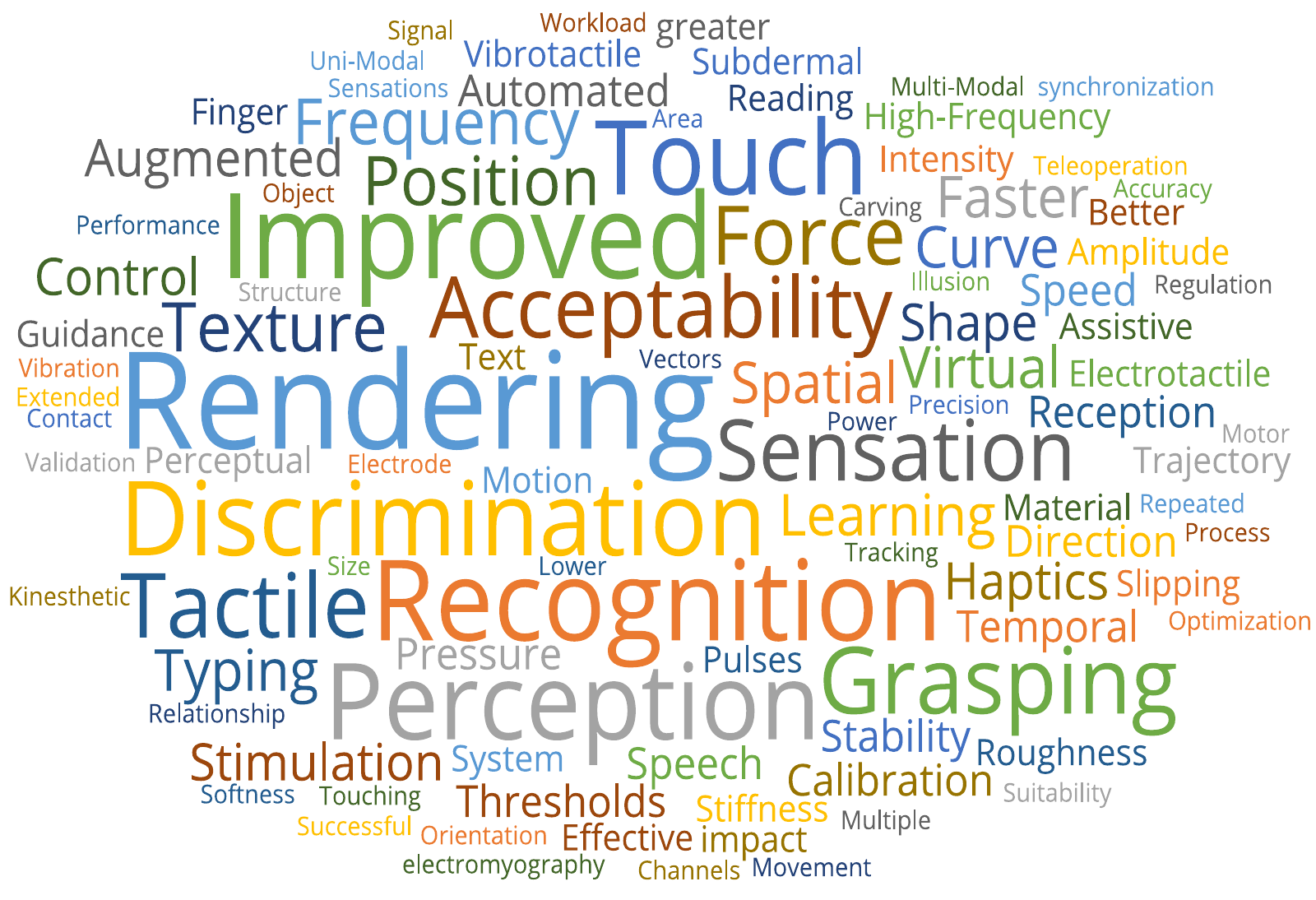}
 \caption{Word cloud: findings frequency.}
 \label{fig:WordCloud}
\end{figure}

\begin{figure}[htbp]
 \centering 
 \includegraphics[width=\columnwidth]{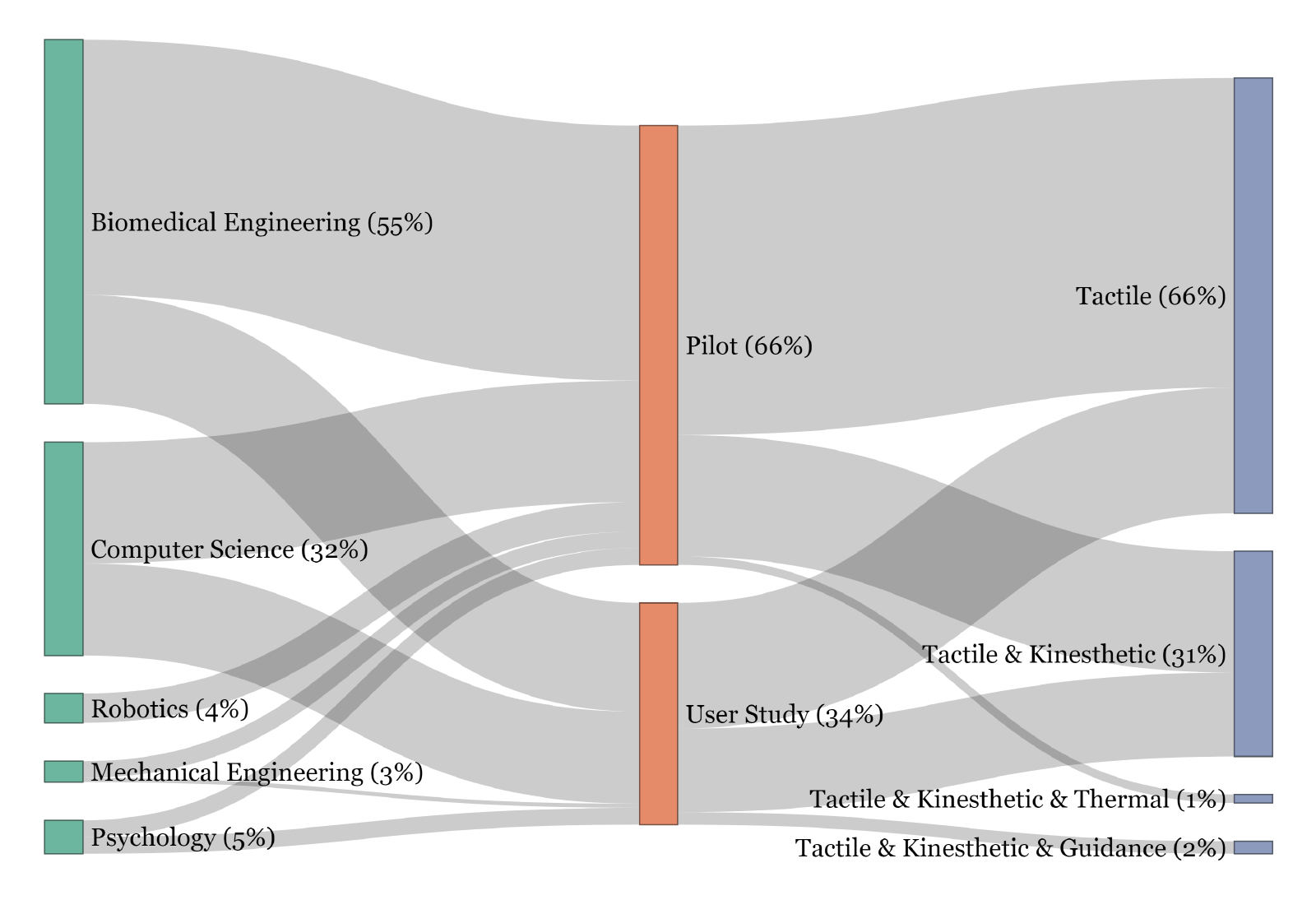}
 \caption{Sanky diagram: discipline, type of study, \& haptic information.\\ Note: The percentages pertaining to electrotactile feedback applications for hand and arm interactions.}
 \label{fig:SankyHI}
\end{figure}

As the \autoref{fig:WordCloud} displays, the most frequent findings pertain to the successful rendering of tactile sensations (e.g., touching a surface), closely followed by the high acceptability of electrotactile feedback by the users. In line with the subsequent most frequent findings, as a result of an efficient touch/contact rendering is the facilitation of equally successful recognition and discrimination of objects based on their shape and size. However, the recognition and discrimination require the elicitation of perceptual processes pertaining to spatial, temporal, textural (e.g., roughness), and motion features. Indeed, as it can be seen in \autoref{fig:WordCloud}, these perceptual processes were also frequently facilitated in the studies that implemented electrotactile feedback. This is also in line with the frequency of eliciting haptic information pertinent to tactile and kinesthetic features (see \autoref{fig:SankyHI}), where the 2/3 offered tactile information, and the 1/3 of the studies offered both tactile and kinesthetic information.%

However, the observed frequencies may have been modulated by the dominance of biomedical engineering applications (see \autoref{fig:SankyHI}), which have predominantly used electrotactile feedback for providing prostheses' users with somatosensory information. Also, this coincides with the particularly high frequency of findings pertaining to performance improvements on force control and grasping (see \autoref{fig:WordCloud}). Nevertheless, the improvements in grasping and force control would not be facilitated without having also benefits on movements' stability, accuracy, and speed, as well as an efficient perception of object's motion, slipping, and stiffness, which can be seen in  \autoref{fig:WordCloud} as frequent findings. These frequent findings indisputably indicate the benefits that electrotactile feedback brings in human-machine/computer-interactions.%

Furthermore, the remaining frequent findings can be seen that they relate to the types of implementations, which were discussed in \autoref{sec:Applications}. Thus, as the \autoref{fig:WordCloud} displays, the electrotactile feedback can be implemented for providing guidance to the user (e.g, in teleoperation or in using portable devices), augmenting haptic sensations, notifying IoT users, improving speech recognition, assisting in virtual typing, and converting conventional text into braille text. However, a downside is that automating and/or optimizing the calibration of electrotactile feedback was scarcely attempted. Furthermore, although that the findings were replicated across diverse disciplines and implementations (see \autoref{fig:SankyHI}, the transferability and applicability of the benefits should be replicated and validated in the field or implementation of interest. Finally, a disadvantage that was observed across the studies, is the sample size, where 2/3 of the studies were pilots (i.e., N $\leq$ 10), while only 1/3 of the studies were user studies with N $>$ 10 (see \autoref{fig:SankyHI}).%

\subsection{Sample Size}
\label{ssec:SampleSize}
Since the sample size defines the generalizability and reliability of the results, the sample size should thus be considered for interpreting responsibly the amassed findings pertinent to the electrotactile feedback. Beyond the dichotomized approach in \autoref{fig:SankyHI}, which showed that 66\% were pilot studies (i.e., N $\leq$ 10), and 34\% were user studies (i.e., N $>$ 10). Only 18\% of the studies had a sample size N $\geq$ 15, and only 6\% had a sample size N $\geq$ 20. This indicates that user studies with a substantial sample size, which examined the utility and potentials of electrotactile feedback, were scarcely conducted in the last decade. Moreover, the 72\% had a sample size N $\leq$ 10. The 41\% of the studies had a sample size N $\geq$ 10, while and 59\% had a sample size N $<$ 10. Notably, only 13\% (i.e., 10/78) recruited patients, while the rest 87\% recruited healthy individuals (predominantly students).%

Although the small sample sizes raise concerns on the generalizability and reliability of the findings, it should be underlined that the findings were replicated across studies and disciplines. The extensive replication of the findings could be attributed to the large effect sizes that were observed in using electrotactile feedback . Nevertheless, the devices were predominantly different, where most of them were not examined in more than one study (i.e., conducting a couple or several follow-up studies), and they were not designed to address comparable user requirements (i.e., substantially different target population). Thus, although the observed advantages of electrotactile feedback in the last decade, the sample size should be considered in future studies.%

\subsection{Stimulation Locations \& Devices}
\label{ssec:LocationDevices}
\begin{figure}[htbp]
 \centering 
 \includegraphics[width=\columnwidth]{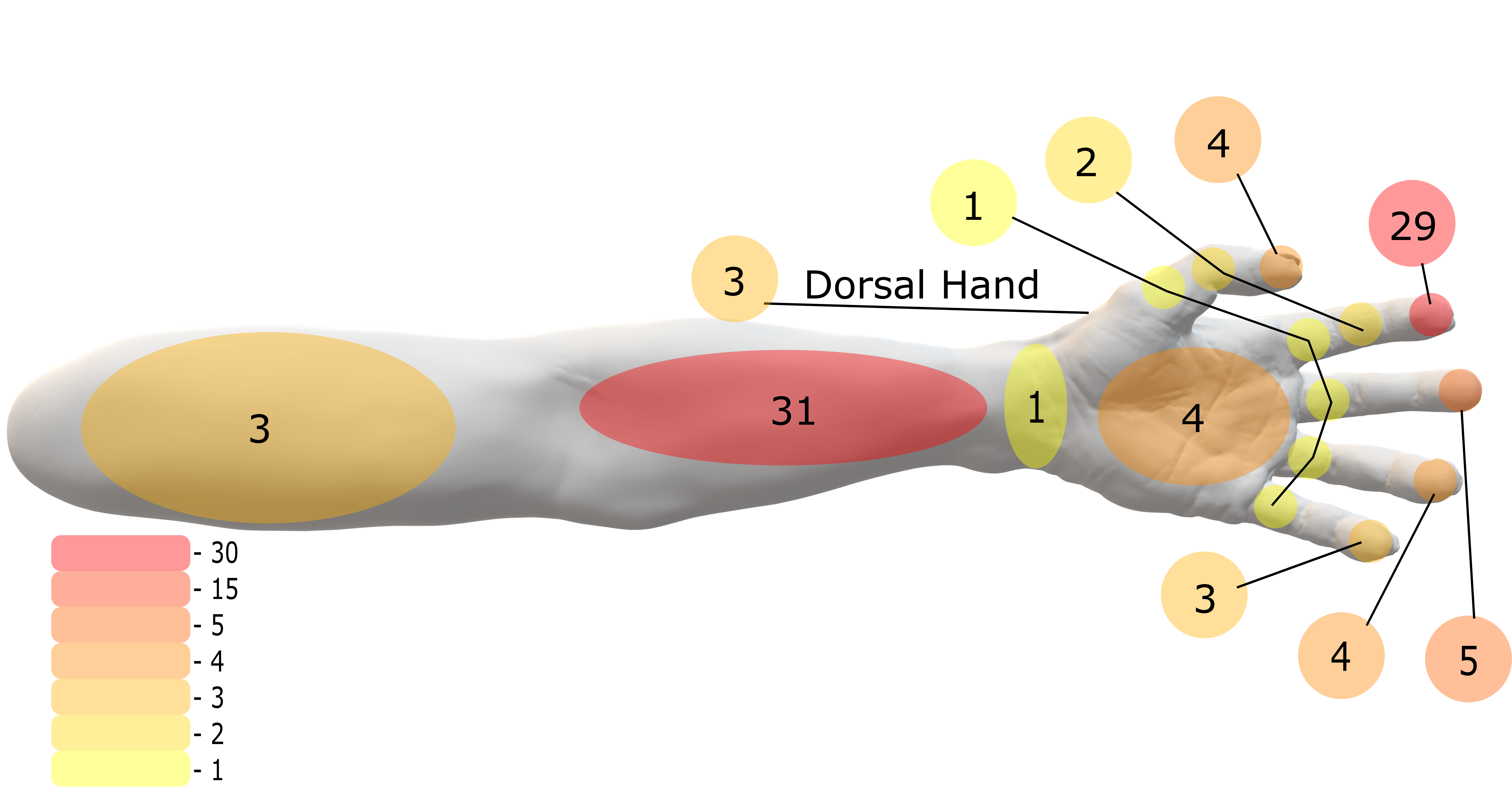}
 \caption{Stimulation Locations' Frequency}
 \label{fig:SLMap}
\end{figure}

\begin{figure}[htbp]
 \centering 
 \includegraphics[width=\columnwidth]{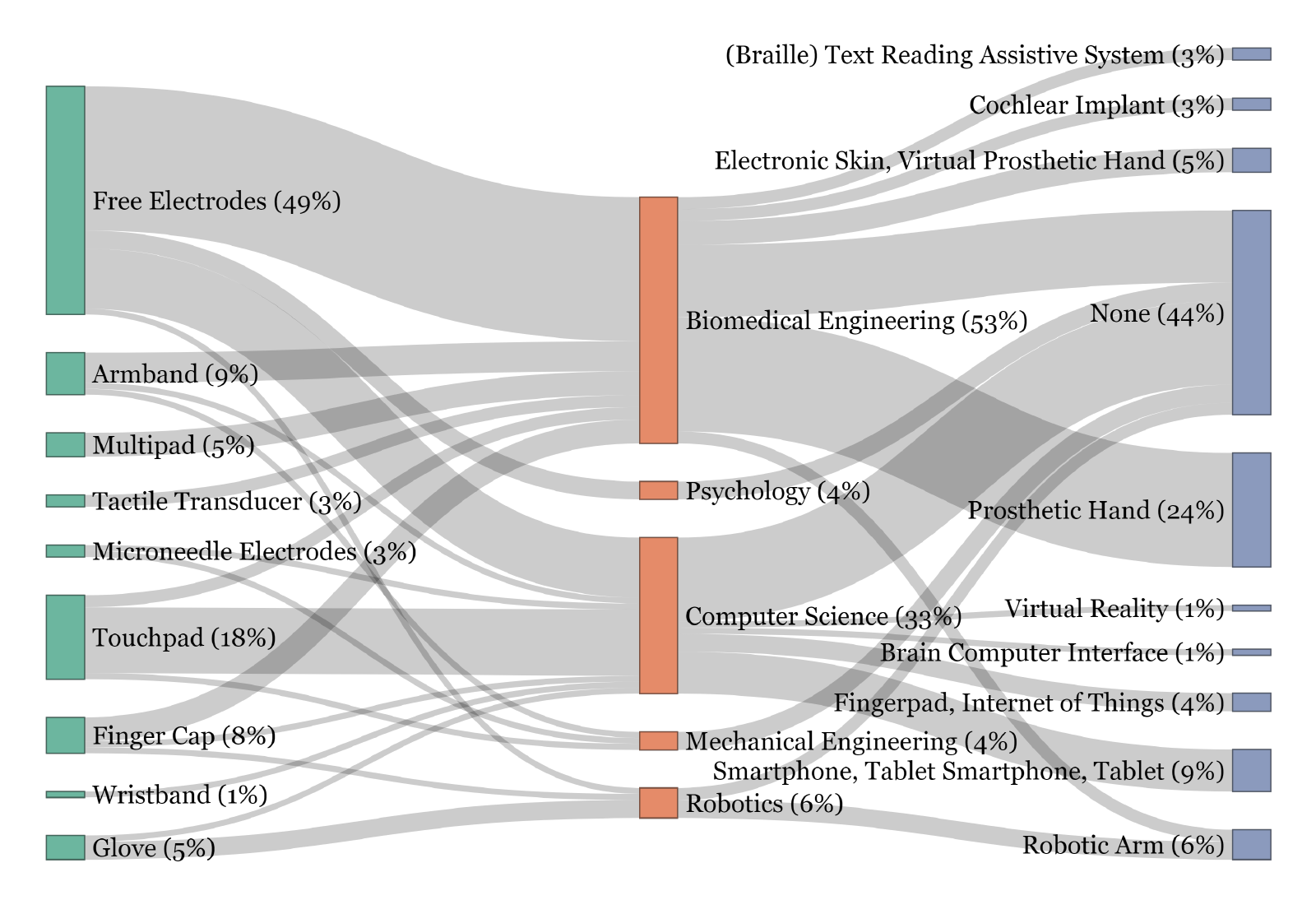}
 \caption{Sanky diagram: haptic device, discipline, \& target device.}
 \label{fig:SankyTD}
\end{figure}

As discussed above, the majority of the implementations of electrotactile feedback were in the field of prosthetics. This can also be seen in \autoref{fig:SLMap}, where 31 studies delivered the stimulation on the forearm and 3 on the upper arm. It should be noted that there are 32 studies that examined electrotaticle feedback for providing prosthetics' users with somatosensory information. Hence, 28 studies stimulated only the forearm, only 1 study the upper arm, and 2 studies stimulated both the forearm and the upper arm. The latter studies were investigating the feasibility to gain comparable benefits by stimulating the upper arm. As it can be seen in \autoref{fig:SankyTD} these studies (i.e., Biomedical Engineering) were using haptic devices such as free electrodes, multipads (i.e., multiple electrodes in a compact design), and armbands (i.e., the stimulator and the electrodes together in a compact device). Also, 24\% of them targeted a prosthetic hand and 5\% of them a virtual prosthetic hand.%

Furthermore, as it was discussed in \autoref{sec:Applications}, the biomedical applications also included a text-reading assistive system (i.e., text to Braille conversion) using a finger cap haptic device, and a speech recognition system using a tactile transducer (see also \autoref{fig:SankyTD}). Thus, the biomedical devices were examined in 37 studies (32 in prosthetics, 3 in Braille, and 2 in Speech recognition) in total. The text-reading assistive and the speech recognition system had integrated an electrotactile feedback, which was delivered to the index fingertip. Thus, from the 29 studies which administered electrotactile feedback on the index fingertip \autoref{fig:SLMap}, 5 of them were for these biomedical applications. As the \autoref{fig:SLMap} illustrates, the 2$^{nd}$ most frequent location for receiving the electrotactile feedback, was the index fingertip. This is a reasonable outcome since the fingertips, especially the index fingertip, are crucial for perceiving tactile and kinesthetic information, which are vital in hand interactions \cite{Edin2008,Johnson2000,Maisto2017,Pacchierotti2017,Westling1987}. In the amassed studies, beyond the aforementioned biomedical applications, the index fingertip was stimulated in human-computer-interaction studies where the user was interacting with portable devices (e.g., tablets and smartphones), virtual environments (i.e., immersive VR), and for teleoperating robotic hands (see \autoref{fig:SankyTD}).%

Unfortunately, the deliverance of electrotactile feedback on the rest of the locations has scarcely been examined. The studies of Pamungkas \emph{et al.} (e.g., \cite{Pamungkas2013,Pamungkas2015b}) were the ones which delivered electrotactile feedback on the dorsal hand (see \autoref{fig:SLMap}). Also, the proximal phalanxes were all stimulated in a single study, where electrotactile feedback was used for virtual typing (see \cite{Pamungkas2019} and \autoref{fig:SLMap}). The wrist was stimulated in a single study, where electrotactile was used for notifying IoT users (see \cite{Stanke2020}). The middle phalanxes were stimulated in studies, where electrotactile feedback was implemented for augmenting haptic sensations (e.g., \cite{Yoshimoto2013,Yoshimoto2015}). The electrotactile feedback on the palm was mainly examined in psychophysical studies (e.g., \cite{Boldt2014,Alotaibi2020}). Equally, the rest of the fingertips were stimulated only in psychophysical studies (e.g., \cite{Kaczmarek2017}) or in teleoperation studies (e.g., \cite{Sagardia2015,Hummel2016}). In summary, as the \autoref{fig:SLMap} indicates, the forearm and the index fingertip were the locations, where the electrotactile feedback was predominantly administered and examined in the studies of the last decade. In contrast, the effects of delivering electrotactile feedback on the rest of the arm's and hand's anatomical locations appear to be under-investigated in this decade.%
\section{Discussion \& Future Directions}
\label{ssec:Discussion}
The systematic review and meta-analysis of the studies which implemented electrotactile feedback for hand interactions in the past decade offered a comprehensive overview of the state-of-art and utility of electrotactile feedback in human-machine/computer-interactions. The electrotactile feedback appeared to elicit tactile sensations and perceptual information, which are essential in hand interactions. Also, the implementation of electrotactile feedback was found to facilitate improvements in performance on various tasks, such as grasping, speech recognition, and virtual typing among others. Furthermore, it was found to be efficient in converting conventional text into Braille, and providing the users with notifications (see \autoref{ssec:GN}), augmented haptic sensations, and guidance. Notably, the advantages of electrotactile feedback were observed and replicated across diverse disciplines and haptic devices, by stimulating different anatomical parts of the arm and the hand. However, the systematic review and meta-analysis have also surfaced and indicated shortcomings and gaps that should be addressed in future studies.%
\subsection{Generalizability \& Reliability of Findings}
\label{ssec:ReliableFindings}
While the amassed findings indicated important benefits of electrotactile feedback for hand interactions, a significant shortcoming was that the studies predominantly had small samples sizes (e.g., N $\leq$ 10). Considering that the implementations target either large populations (e.g., users of portable devices, wearable devices like smartwatches, and immersive VR), or clinical populations (e.g., amputees, legally blind, and individuals with hearing impairments), the sample size is even more important for reliably generalizing the observed findings.%
 Under-powered statistical analyses may inflate or deflate considerably the observed effect sizes \cite{Ioannidis2008}. In human-machine/computer-interaction studies, the effect sizes are of utter importance for examining the utility of the medium or technique\cite{Dragicevic2016}. To address the sample size issues, future studies should strive to examine the effects and benefits in studies with larger sample sizes. We suggest a minimum of N = 10 for preliminary results, and N $\geq$ 15 for comprehensive user studies, especially for the studies which explore biomedical applications. Importantly, we suggest the examination of effect sizes to be conducted by using parsimonious metrics such as the $Hedges'$ $g$ (for t-tests and comparisons) and the $\omega^2$ (for ANOVA and regression analyses)\cite{Lakens2013}.%
\subsection{Technical Improvements}
\label{ssec:TechImprov}
As discussed in \autoref{sec:Overview}, a substantial disadvantage of electrotactile feedback is that may cause skin irritation, burns, electric shock, and discomfort ~\cite{Kaczmarek1991}. To prevent such undesirable effects, a calibration may be conducted to adjust the intensity and duration of the electrotactile feedback to the personal sensational and discomfort thresholds of each individual/user~\cite{Stephens2018}. However, physiological changes like perspiration may fluctuate skin's conductivity, which consequently may cause a lack of sensation or discomfort respectively to the fluctuations \cite{Bach2010}. While it was considered that the sensation and discomfort thresholds predominantly increase across the duration of exposure to electrotactile feedback, the study of Vizcay \emph{et al.} \cite{Vizcay2021} showed that after familiarization with the electrotactile feedback, then the sensation and discomfort threshold does not appear to significantly change. However, the absence of significant changes was observed after the 2$^{nd}$ calibration, which indicates that this is a product of increased tolerance and/or stabilization of skin's conductivity \cite{Vizcay2021}.%

Furthermore, the skin's motion due to applied force for the contact, may cause substantial alterations to the perceived intensity of electrotactile feedback \cite{Kajimoto2012b}. This shortcoming remains an open issue. However, this limitation applies only to applications of electrotactile feedback, where the contact with physical objects (e.g., tablets, smartphones, or tangible objects) is direct. In indirect (e.g., prosthetics) or mid-air (e.g., in VR) interactions, this limitation of electrotactile feedback does not appear as an issue since there is not a force applied to the location of the electrode. Another factor that should be considered is the location of the stimulation (i.e., diverse skin's thickness) which requires diverse sensation and discomfort threshold proportionally ~\cite{Stephens2018}. Therefore, the calibration should be done for each location of stimulation, and it has also to be performed recurrently for readjusting the sensation and discomfort thresholds proportionally to the perceptual changes (e.g., tolerance) and changes of skin's conductivity.%

Several studies have strived to automate and optimize the calibration process. Gregory \emph{et al.} \cite{Gregory2009} delivered the Cole-Cole circuit model for identifying and characterizing the parameters of the skin-electrode interface. Similarly, Fadali \emph{et al.} \cite{Fadali2009} used a Takagi-Sugeno-Kang (TSK) fuzzy logic modelling approach for identifying tactile preferences. Kajimoto \cite{Kajimoto2012b} developed a real-time feedback loop, where the skin's impedance was monitored, and the electrotactile feedback's intensity was modulated respectively. Similarly, the studies of Akhtar \emph{et al.} \cite{Akhtar2014,Akhtar2018} offered models measuring the impedance of electrode-skin interface to reduce the variability and optimize the electrotactile feedback parameters. The studies of Rahimi et al \cite{Rahimi2019,Rahimi2019b} attempted to provide an automated recurrent calibration process for the users (legally blind) of a Braille text-reading assistive system. Furthermore, Isakovic \emph{et al.} \cite{Isakovic2019} endeavoured optimization of the calibration process in order to provide the users with a personalized electrotactile feedback. Although that these are studies in the desired direction, further studies are required for achieving an optimal, automated, and recurrent calibration. Given its importance in user's experience and performance, this should be prioritized in future studies.%

Moreover, the material of the electrodes is equally important for delivering pleasant and efficient electrotactile feedback ~\cite{Kaczmarek1991}. A recent review suggested that the material of contemporary electrodes meets these criteria~\cite{Jung2020}. However, as it was also seen in Withana \emph{et al.} \cite{Withana2018} study, while these electrodes are light, flexible, efficient in eliciting tactile sensations, their main disadvantage is their limited durability. Similarly, in the field of prosthetics, the studies of Isakovic \emph{et al.} \cite{Isakovic2016} and Strbac, \emph{et al.}, \cite{Strbac2016} implemented flexible electrodes integrated with a multichannel stimulator. However, the electrodes in these studies were also disposable with limited durability. On a positive note, the recent studies of Stephens \emph{et al.} \cite{Stephens2018b,Stephens2020} showed that concentric electrodes coated with a conductive graphene ink are equally flexible and efficient in delivering tactile sensations, and, importantly, are reusable with estimated durability over a year. However, more research is required for elucidating these promising findings of concentric graphene electrodes.%

\subsection{Wearable \& Portable Devices}
\label{ssec:WearablePortable}
As discussed above, contemporary haptic devices should also be wearable and/or portable. The wearability of haptic devices not only improve their overall efficiency, but they also enable their frequent use in everyday life \cite{Pacchierotti2017,Maisto2017,Meli2018,Chinello2019}. The electrotactile feedback was predominantly integrated into devices with low wearability (e.g., see (d) in \autoref{fig:HD}), although that they reported advantages of using electrotactile feedback. However, other studies like Cheng and Zhang \cite{Cheng2017}, Withana \emph{et al.} \cite{Withana2018}, Strbac \emph{et al.} \cite{Strbac2019}, presented and implemented haptic devices, which are lightweight and wearable in a compact design (see (b), (e), and (f) in \autoref{fig:HD}). Despite that the two devices are destined for prosthetics and the other one for augmenting haptic sensations, they provide evidence in support of the wearability of electrotactile interfaces. Furthermore, the recent study of Stanke \emph{et al.} \cite{Stanke2020}, by using a wristband, provided IoT users with notifications. Given that contemporary wearable devices are frequently worn on the wrist (e.g., \emph{Apple Watch}) \cite{Pacchierotti2017}, the study of Stanke \emph{et al.} \cite{Stanke2020} provides evidence that the electrotactile feedback can be feasibly integrated with such everyday devices.%

Furthermore, the electrotactile feedback was efficiently delivered through transparent haptic touchpads on portable devices such as smartphones and tablets. The \emph{Skeletouch} of Kajimoto \cite{Kajimoto2012} on a smartphone's touchscreen and the \emph{E-Pad} (see (a) in \autoref{fig:HD}) of Wang \emph{et al.} \cite{Zhao2018} on tablet's touchscreen are the two most representative examples. Considering that tablets and smartphones have become indispensable tools in everyday life, which are used to alleviate the daily cognitive workload and improve performance and productivity on everyday tasks and activities \cite{Carter2016,Wilmer2017}, the use of haptic sensation may improve the performance and usability of these devices. Interestingly, the study of Wang \emph{et al.} \cite{Zhao2018} showed that electrotactile feedback may provide the user with guidance and improve spatial performance. Nevertheless, none of the electrotactile feedback systems was integrated with the portable device in a compact design. Future iterations should attempt to derive with a compact and practical design of electrotactile systems integrated with portable devices. Finally, the majority of studies predominantly stimulated the index fingertip. Future studies should also attempt to examine the effects of multi-finger stimulation in corresponding tasks which require interactions using multiple fingers.%
\subsection{Prostheses}
\label{ssec:DiscProstheses}
As discussed in \autoref{sec:Applications} and \autoref{sec:Meta}, approximately half of the implementations of electrotactile feedback were in prosthetic hands' users. They produced robust evidence regarding the advantages of electrotactile feedback for providing the users with somatosensory information, especially in closed-loop systems with EMG. Electrotactile feedback was found to efficiently render tactile sensations, and eliciting perceptual processes pertaining to spatial (e.g., shape and size), textural (e.g., softness), and kinesthetic (e.g., object's or hand's motion) properties. Importantly, this enriched somatosensory information resulted in performance improvements (e.g., force control, grasping and object's manipulation accuracy), as well as in an improved acceptance and embodiment of the prostheses.%

However, only 3 out of 32 studies (i.e., 9\%) examined electrotactile stimulation on the upper arm. Examining the administration of electrotactile feedback on the upper arm would facilitate the replication of the findings (i.e., comparable to forearm stimulation) and postulate the applicability and utility of electrotactile feedback in prostheses (e.g., in closed-loop systems) for individuals which were amputated at a point on the forearm or elbow. The studies which have done so (e.g., Choi \emph{et al.} \cite{Choi2017} and Cheng \emph{et al.} \cite{Cheng2019}) produced findings postulating that electrotactile feedback on the upper arm presents advantages comparable to forearm's stimulation. Nevertheless, future studies should strive to investigate further the administration of electrotactile feedback on the upper arm.%

Furthermore, only 7 out of 32 studies (i.e., 22\%) examined these advantages in patients (i.e., amputees). Considering that the final aim is the amelioration of acceptance and performance of prostheses, the evaluation of the systems integrating electrotactile feedback in end-users (i.e., amputees) is of utter importance. Hence, prospective studies should attempt to examine the proposed systems in clinical populations. However, the proposed systems should be integrated with products (i.e., prosthetic hands or arms) that aim to be released in the market, in order to elucidate on the benefits of their system. In this systematic review, only one study was identified that conducted such a study. Strbac \emph{et al.},  \cite{Strbac2017} integrated their closed-loop system with \emph{Michelangelo hand} (a prosthetic hand which is already in the market) for evaluating its performance. Future studies should strive to evaluate the proposed systems by integrating them with prosthetics that are either already or intending to be released in the market.%

Finally, as mentioned above, the electrotactile was used with EMG in closed-loop systems. Alternatively, the study of Achanccaray and Hayashibe \cite{Achanccaray2020} integrated electrotactile feedback with BCI,  where improved motor imagery and significant improvements in flexion and extension were observed. Given that flexion and extension are the main movements in grasping, the results of this study indicate that electrotactile with BCI systems may show benefits comparable to closed-loop systems. Future studies may attempt to examine further the BCI with electrotactile combination, which may be found to be an efficient alternative to closed-loop systems.%

\subsection{Teleoperation \& Virtual Reality}
\label{ssec:DiscTeleVR}
As discussed above, prosthetic hands' users were provided with somatosensory substitution by implementing electrotactile interfaces. This somatosensory substitution facilitated an embodied perception of the prostheses, which also assisted the users with improving their performance on tasks requiring grasping and handling items. Similarly, telepresence, which takes place in teleoperation systems, requires a sense of embodiment for the operated tool (e.g., robotic hand) which facilitates an improved control and performance. Indeed,  comparable performance benefits were observed on teleoperation tasks requiring handling items (e.g., \cite{Sagardia2015,Hummel2016,Meli2014}). Also, the majority of the teleoperation systems involved immersive VR (e.g., \cite{Pamungkas2013,Pamungkas2015,Sagardia2015,Hummel2016}). This was well expected since the telepresence in teleoperation systems is comparable to the sense of presence in immersive VR.%

In immersive VR, the sense of presence depends on the strength of the placement (i.e., realistic visuals), plausibility (i.e., realistic interactions), and embodiment illusions \cite{Kourtesis2020,Slater2009,Slater2010}. However, the placement illusion is fragile, especially when the virtual environment does not realistically react to the user's actions \cite{Slater2009,Slater2010}. This is resolved by the plausibility illusion, which is the deception of the user that the environment reacts to the user's actions. Notably, the strength of the plausibility illusion is analogous to the sensorimotor contingency, which is the integration of the senses such as vision, hearing, touch, kinesthesia, smell, and taste \cite{Kourtesis2020,Gonzalez2017}. The haptic feedback pertains both to the sense of touch and kinesthesia (i.e., proprioception) \cite{Argelaguet2016,Kourtesis2020,Prattichizzo2012}. The haptics therefore additionally reinforce the plausibility illusion by providing expected haptic feedback to the user \cite{Argelaguet2016,Kourtesis2020}. Also, the haptic feedback is central in embodiment illusion \cite{Argelaguet2016,Gonzalez2019,Peck2021}. There is multidirectional interaction between haptic sensations and embodiment illusion, where the haptic feedback appears to reinforce the embodiment illusion \cite{Peck2021}, while a strong embodiment illusion (i.e., virtual body ownership) appears to intensify the perceived strength of the haptic sensation \cite{Gonzalez2019}.%

However, the wearability/portability of the haptic device is vital in immersive VR applications \cite{Pacchierotti2017,Maisto2017}. As discussed above, the electrotactile seems to facilitate a compact and wearable design, which can be implemented in immersive VR applications for providing the users with haptic information \cite{Jung2020}. Nevertheless, in this systematic review, only 6 studies used immersive VR, where most of the time was used for teleoperation purposes (e.g., Sagardia \emph{et al.} \cite{Sagardia2015}) and frequently had very small sample sizes (e.g., Pamungkas \emph{et al.} \cite{Pamungkas2013,Pamungkas2015}). Of course, immersive VR is intrinsically related to teleoperation, since it offers improved telepresence to the operator \cite{Almeida2017}. Thus, it does not come with surprise the frequency of implementing immersive VR for teleoperation purposes. Given that there are common requirements in prosthetics (i.e., embodiment), teleoperation (i.e., telepresence and embodiment), and immersive VR (i.e., sense of presence and embodiment), it would be expected that electrotactile feedback would facilitate improvements in user experience (i.e., presence) and performance. Thus, it was unexpected that a scarcity of electrotactile feedback implementations in teleoperation and immersive VR, especially in the latter, will be observed in the last decade.%

The studies which implemented electrotactile feedback for improving immersive VR experiences, using VR head-mounted displays, were the small pilot study of Pamungkas and Ward \cite{Pamungkas2016}, and the pilot study of Withana \emph{et al.} \cite{Withana2018}. Thus, it cannot be robustly supported that the improvements in embodiment and performance, which were observed in prosthetics' users, will be replicated in immersive VR. However, regarding performance, the user study of Hummel \emph{et al.} \cite{Hummel2016}, using a VR CAVE system, showed that electrotactile facilitates a better grasping. Also, the recent user study of Vizcay \emph{et al.} \cite{Vizcay2021}, using a VR head-mounted display, equally reported improvements in contact accuracy and precision. Although that these studies provided evidence in an adequate sample size (N = 19 and 21 respectively), further research is required for elucidating on the performance gains in immersive VR by administering electrotactile feedback. Importantly, no study has investigated the impact of electrotactile feedback on presence and embodiment in immersive VR. Considering that the sense of presence and embodiment are the cornerstones of immersive VR, future studies should hence endeavour to scrutinize the contribution of electrotactile feedback in them.%

\section{Synopsis \& Conclusions}
\label{sec:Conclusion}
In this paper, we have conducted a systematic review of the electrotactile feedback implementations for various hand interactions in human-computer/machine-interaction systems. The concentrated studies were discussed per type of stimulation or implementation. Furthermore, a meta-analysis of the studies was additionally performed, which highlighted the replicability of the findings across diverse electrotactile feedback systems, as well as the methodological drawbacks and knowledge gaps. Finally, a comprehensive discussion of all the above was attempted to offer perspectives on the current state-of-art and future directions.%

The amassed studies indicated that electrotactile feedback was predominantly used for biomedical applications, especially for providing prosthetic hands' users with somatosensory information. The early attempts were made using free electrodes attached either (principally) on the forearm or the upper arm. More recent attempts involved a more ergonomic armband that includes both the electrodes (e.g., in a multipad form) and the stimulator. Other biomedical applications pertained to cochlear implant users and readers of Braille texts. Furthermore, electrotactile feedback was implemented to facilitate successful robotic teleoperation and human-computer interaction in VR. Finally, haptic interfaces were used for providing users of portable devices (e.g., smartphones and tablets) with electrotactile sensations, as well as for augmenting haptic information.%

Although the electrotactile feedback was predominantly used in biomedical engineering (approximately half of the applications), comparable findings were found to every type of application for human-machine/computer-interaction regardless of the discipline and the location of stimulation. This interdisciplinary replication of the amassed findings allows their presentation in a unifying manner. Hence, one could conclude that electrotactile feedback was found to be well accepted by diverse users. Also, it was successful in rendering and/or augmenting several tactile sensations pertinent to touch and textures. Electrotactile feedback elicited multiple perceptual processes such as spatial, temporal, kinesthetic (e.g., motion, force, and slipping), and textural (e.g., roughness) perception. The haptic information provided by electrotactile feedback assisted the users to have a greater force control and grasping performance, which further facilitated effective teleoperation of robotic hands and tools. Furthermore, electrotactile feeling was also efficient in providing guidance and improving speech reception, as well as facilitating Braille text reading. Finally, it was found efficient in notifying the user by eliciting a state of alertness, as well as for guiding the user in performing fine-movement tasks such as carving and virtual mid-air typing.%

Although a recurrent limitation pertaining to small sample sizes using diverse actuators/stimulators was observed and discussed, the findings were replicated regardless of the location of the stimulation and the implemented device. Nevertheless, future studies should strive to examine the usability of electrotactile feedback in larger sample sizes and conduct appropriate statistical analyses respectively. A priority should be the automation and optimization of electrotactile feedback's recurrent calibration, as well as improved durability of electrodes. Of course, the wearability and/or portability of electrotactile feedback should be another future aim by integrating it in devices such as smartphones, tablets, and smartwatches. Also, electrotactile feedback systems may be integrated with commercial prosthetic hands/arms, for providing sensory substitution by stimulating the upper-arm or forearm, where the long-term effects on the users' performance, acceptability, and embodiment may be scrutinized. Finally, the same benefits on performance and embodiment may also be investigated in immersive VR and teleoperation systems, where these factors are crucial for the systems' effectiveness and user's experience.%
\ifCLASSOPTIONcompsoc
  \section*{Acknowledgments}
\else
  \section*{Acknowledgment}
\fi
This work was supported by the European Union's Horizon 2020 research and innovation program under grant agreement No. 856718 (TACTILITY).
\bibliographystyle{IEEEtran}
\bibliography{my.bib}
\begin{IEEEbiography}
[{\includegraphics[width=1in,height=1.25in,clip,keepaspectratio]{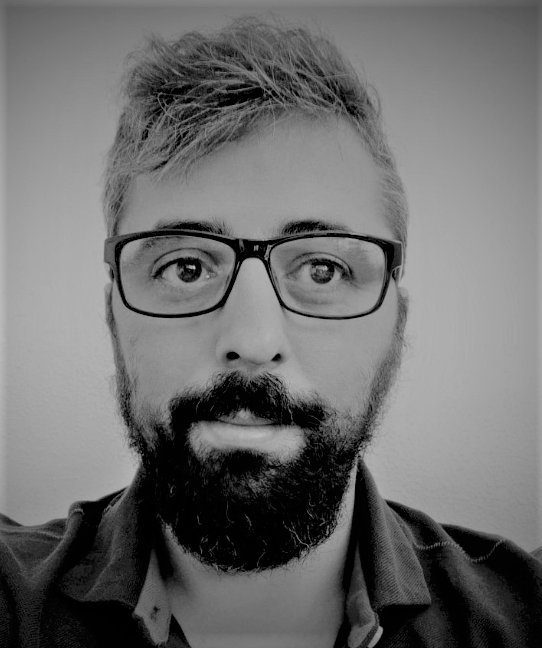}}]{Panagiotis Kourtesis} is a postdoctoral researcher of the National Research Institute of France for Computer Science and Automation (Inria). 
Since 2020, he works on the TACTILITY project, where he investigates the human factors, perception, and motor performance pertaining to the use of electrotactile feedback in VR. 
In 2020, he obtained his PhD in Experimental Psychology and Cognitive Neuroscience at the University of Edinburgh, United Kingdom,  where he explored the use of VR methods in psychological sciences.%
\end{IEEEbiography}
\begin{IEEEbiography}
[{\includegraphics[width=1in,height=1.25in,clip,keepaspectratio]{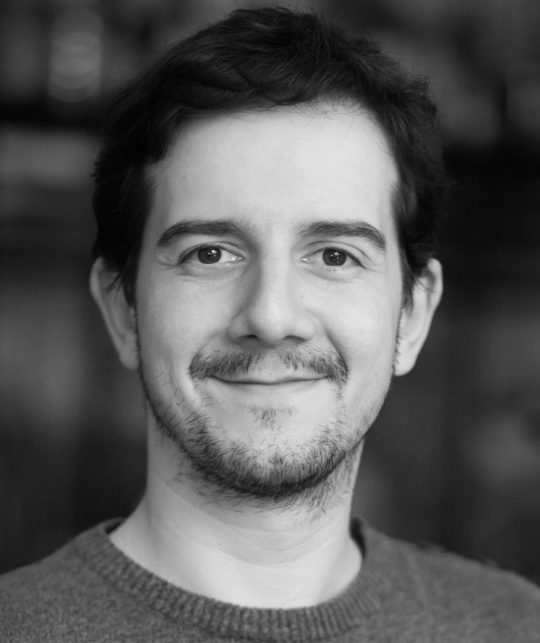}}]{Ferran Argelaguet} is a Research Scientist at Inria, the French research institute in computer science and applied mathematics since 2016. He received his Habilitation in Computer Science from the University of Rennes 1 in 2021 (France), and his Ph.D. in Computer Science from the Universitat Politècnica de Catalunya in 2011 (Spain). His research interests include virtual and augmented reality, human perception and cognition, and human-computer interaction.%
\end{IEEEbiography}
\begin{IEEEbiography}
[{\includegraphics[width=1in,height=1.25in,clip,keepaspectratio]{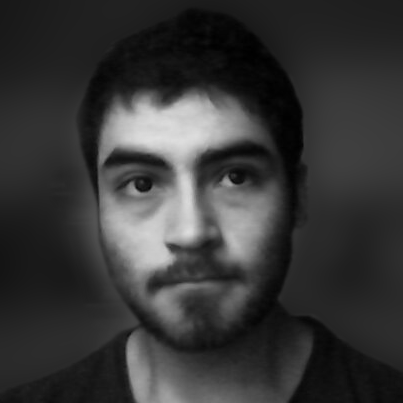}}]{Sebastian Vizcay} is a PhD student at Inria Rennes (France). He earned his BE and ME from the University of Santiago (Chile) in 2013 and 2015, respectively. He was a research intern at the University of Ottawa (Canada) in 2014. He previously worked as computer graphics engineer at Inria Sophia-Antipolis (France) for the EU Horizon 2020 project EMOTIVE. Since 2019 he is working towards his PhD in Electrotactile Feedback in Virtual Reality in the scope of the EU Horizon 2020 project TACTILITY.%
\end{IEEEbiography}
\begin{IEEEbiography}
[{\includegraphics[width=1in,height=1.25in,clip,keepaspectratio]{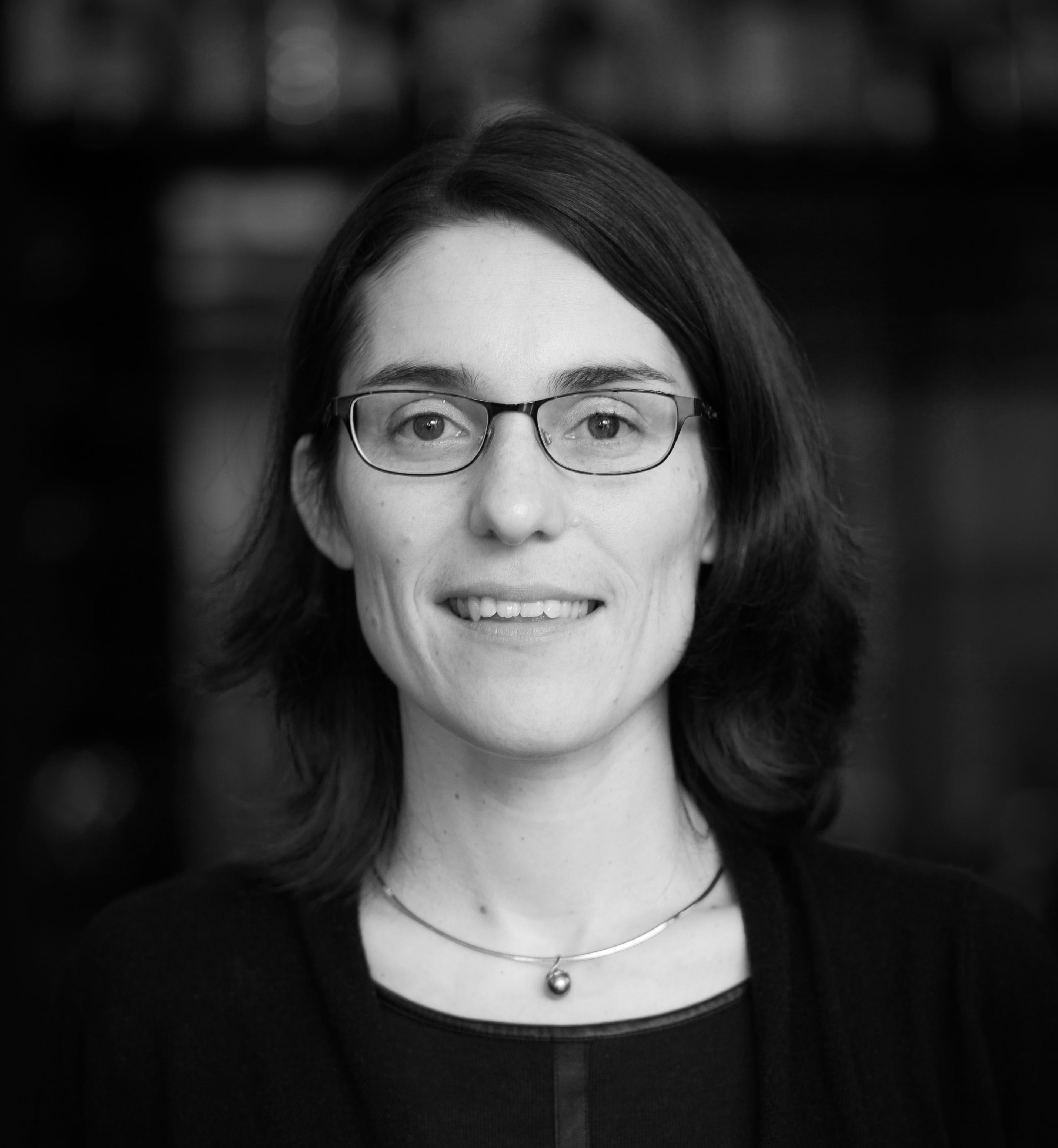}}]{Maud Marchal} is a Full Professor in Computer Science at Univ. Rennes, INSA/IRISA, France. She received her Habilitation from the University of Rennes 1 in 2014 and her PhD from University Joseph Fourier in Grenoble in 2006. Her research interests include haptics, virtual reality and 3D interaction, physics-based simulation. She is Associate Editor of IEEE Transactions on Haptics since 2021.%
\end{IEEEbiography}
\begin{IEEEbiography}[{\includegraphics[width=1in,height=1.25in,clip,keepaspectratio]{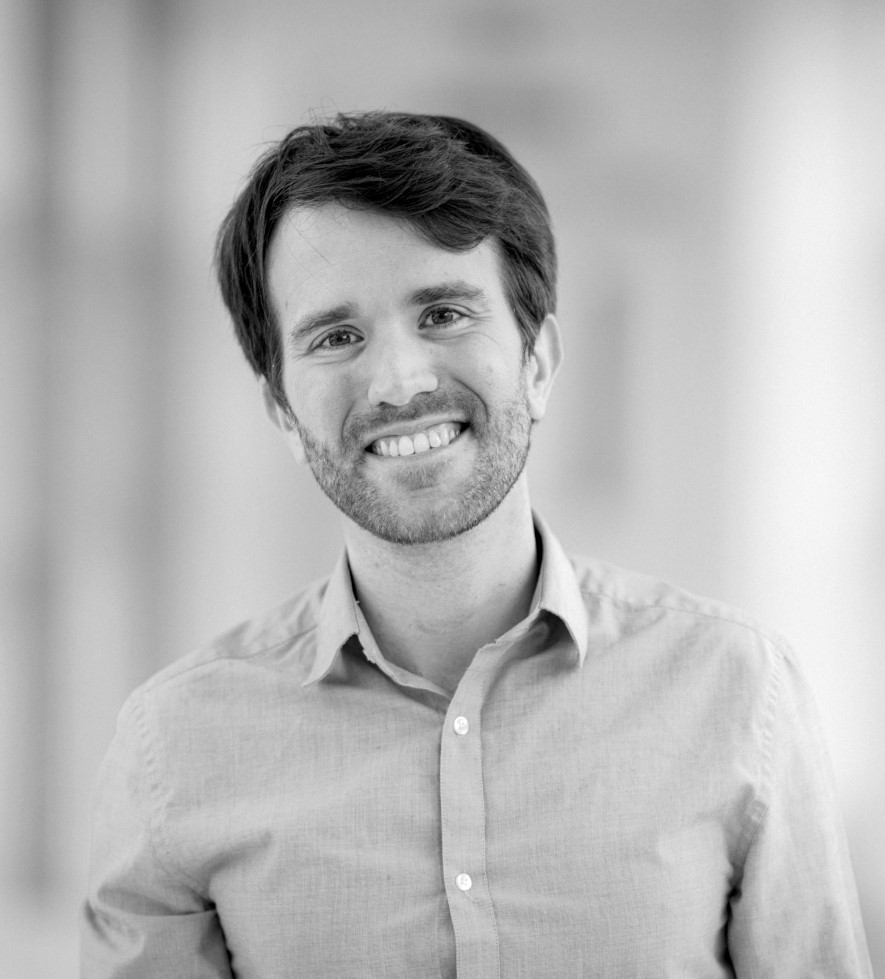}}]{Claudio  Pacchierotti} (S'12, M'15, SM'20) is a tenured researcher of the CNRS in Rennes, France.
He was previously a postdoctoral researcher of the Italian Institute of Technology, Genova, Italy.
Pacchierotti earned his Ph.D. degree at the University of Siena in 2014 and his HDR at the University of Rennes 1 in 2022.
He visited the Penn Haptics Group at the University of Pennsylvania in 2014, the Institute for Biomedical Technology and Technical Medicine (MIRA) of the University of Twente in 2014, and the Department of Computer, Control and Management Engineering of Sapienza University of Rome in 2022.
Pacchierotti received the 2014 EuroHaptics Best PhD Thesis Award for the best doctoral thesis in the field of haptics.
He has also been an Associate Editor for various haptic conferences as well as member of the Conference Editorial Board for Eurohaptics 2022. He is Senior Chair of the IEEE Technical Committee on Haptics and Secretary of the Eurohaptics Society.
\end{IEEEbiography}
\vfill
\end{document}